\documentclass[english,aps,prc,nofootinbib,superscriptaddress,twocolumn,letterpaper,floatfix]{revtex4-1}
\usepackage[utf8]{inputenc}
\usepackage{hyperref}
\hypersetup{
  colorlinks=true,
  linkcolor=blue,
  citecolor=cyan,
  urlcolor=blue
}
\usepackage{breakurl}

\usepackage{graphicx}
\usepackage{amssymb,amsmath}
\usepackage{bm}
\usepackage[usenames]{color}
\usepackage{times}
\usepackage{comment}

\newcommand{\trento}{T\raisebox{-0.5ex}{R}ENTo}
\newcommand{\NA}{N_{\text{part}}^A}
\newcommand{\NB}{N_{\text{part}}^B}
\newcommand{\Npart}{N_{\text{part}}}

\allowdisplaybreaks

\begin{document}

\preprint{This line only printed with preprint option}

\title{Initial spin fluctuations as a probe of cluster spin structure in $^{16}\mathrm{O}$ and $^{20}\mathrm{Ne}$ nuclei}

\author{Xiang Fan}

\affiliation{Institute of Particle Physics and Key Laboratory of Quark and Lepton 
Physics (MOE), Central China Normal University, Wuhan, Hubei, 430079, China}

\author{Jun-Qi Tao}
\affiliation{Institute of Particle Physics and Key Laboratory of Quark and Lepton 
Physics (MOE), Central China Normal University, Wuhan, Hubei, 430079, China}
\affiliation{School of Science and Engineering, The Chinese University of Hong Kong, Shenzhen (CUHK-Shenzhen), Guangdong, 518172, China}

\author{Ze-Fang Jiang}
\email{jiangzf@mails.ccnu.edu.cn}
\affiliation{Department of Physics and Electronic-Information Engineering, 
Hubei Engineering University, Xiaogan, Hubei, 432000, China}

\author{Ben-Wei Zhang}
\email{bwzhang@mail.ccnu.edu.cn}
\affiliation{Institute of Particle Physics and Key Laboratory of Quark and Lepton 
Physics (MOE), Central China Normal University, Wuhan, Hubei, 430079, China}

\date{\today}

\begin{abstract}
We investigate the imprint of $\alpha$ clustering on initial spin fluctuations in relativistic $^{16}\mathrm{O}+{}^{16}\mathrm{O}$ and $^{20}\mathrm{Ne}+{}^{20}\mathrm{Ne}$ collisions at $\sqrt{s_{\mathrm{NN}}}=5.36$~TeV.
Utilizing \textit{ab initio} configurations from Nuclear Lattice Effective Field Theory (NLEFT) and phenomenological $\alpha$-cluster models within the \trento{}  framework, 
we compute the event-by-event variance of the initial net spin polarization. 
We find that the strong short-range spin--isospin correlations characteristic of $\alpha$ clusters lead to a significant suppression of spin fluctuations compared to a spherical Woods--Saxon reference without additional intrinsic spin correlations beyond the global $J=0$ constraint.
By constructing a scaled fluctuation observable that accounts for the dominant trivial finite-size effect, we demonstrate that this suppression exhibits a non-monotonic centrality dependence sensitive to the detailed cluster geometry.
Furthermore, we propose the ratio of scaled spin fluctuations between $^{20}\mathrm{Ne}$ and $^{16}\mathrm{O}$ systems as a potentially discriminating quantity. Our results predict distinct percent-level deviations from the baseline for clustered nuclei,
suggesting that measurements of final-state $\Lambda$-hyperon spin correlations may provide new constraints on the ground-state spin structure of light nuclei.
\end{abstract}

\maketitle

\section{Introduction}\label{sec:intro}

Relativistic heavy-ion collisions at RHIC and the LHC create strongly
interacting QCD matter at extreme temperature, vorticity, and magnetic fields,
commonly referred to as the quark--gluon plasma
(QGP)~\cite{Adcox:2005,Muller:2012,Shou2024QCDMatterALICE,Ma2026QuarkSoupTemperature}. The observation
of a global $\Lambda$-hyperon polarization in non-central heavy-ion
collisions~\cite{Adamczyk:2017,Acharya:2020} -- in agreement with early
theoretical predictions of spin--orbit coupling in the
QGP~\cite{Liang:2005,Gao:2007bc,Becattini:2008} -- has opened a new avenue
to study the spin degrees of freedom of the QGP at high
energies~\cite{STAR:2022fan,Chen:2024afy}. The overall magnitude and
beam-energy dependence of the global $\Lambda$ polarization are reasonably
described by several hydrodynamic and transport
models~\cite{Becattini:2017,Xie:2017,Fu:2020oxj,
Li:2021_MultiStrangeFeeddown,Chen:2024aom,Deng2020VorticityLowEnergyHIC}, improving our understanding of
spin--orbit coupling and the vortical nature of the QGP fluid. At the same
time, the so-called ``sign puzzle'' in local $\Lambda$ polarization
measurements -- where the measured azimuthal dependence of the longitudinal
polarization is opposite to theoretical expectations -- has prompted renewed
discussions about the origin and transport of spin in the
QGP~\cite{Adam:2019,Becattini:2018,Xia:2019,Wu:2019eyi,Yi:2021ryh}.

More recently, attention has shifted from event-averaged polarization to spin
fluctuations and spin
correlations~\cite{Baumgart:2012ay,Fucilla:2025kit,Afik:2025grr,Han:2024ugl,Liang2026QCDConfinementSpinSpin}.
On the theory side, hyperon-pair spin correlators have been formulated
as sensitive probes of spin dynamics in the QGP, hyperon-pair production
mechanisms, initial spin-density fluctuations, and quantum spin
correlations in the hyperon sector~\cite{Giacalone:2025,Pang:2016,
Lyuboshitz:2010,Ellis:2011,Lin:2025}.
Experimentally, first measurements of hyperon pair spin correlations at RHIC
demonstrate that such observables are accessible with good precision and
already begin to constrain scenarios for spin transport and nontrivial quantum
correlations in QCD matter~\cite{STAR:2025njp}. More broadly, polarization
observables of hypernuclei have been proposed as tools to decipher their
internal spin structure in heavy-ion collisions, e.g.\ for
(anti)hypertriton~\cite{Sun:2025oib}.

At the same time, collisions of a growing variety of nuclear species
(including $^{238}$U, $^{129}$Xe, $^{96}$Ru, $^{96}$Zr, $^{208}$Pb,
$^{20}$Ne, and $^{16}$O) have opened up the possibility of imaging nuclear
structure with high-energy data~\cite{Jia:2025wey,
Jia2024ImagingInitialCondition}, including nuclear deformation,
neutron-skin effects, and cluster structures~\cite{STAR:2024wgy,
Giacalone:2023PRL,Bally:2022PRL,Liu:2025Ne20Alpha,Ding:2023O16Signals,
Zhang:2024vkh,Giacalone:2024ixe,Giacalone2024BeyondAxialSymmetry,
Schenke2024HexadecapoleDeformation,Ding2024NeutronSkin}. In particular, 
light self-conjugate $N=Z$ nuclei such as $^{16}$O and
$^{20}$Ne are known to exhibit pronounced $\alpha$-cluster structures, a
phenomenon well-established by a long history of low-energy nuclear
experiments, phenomenological studies, and recent
reviews~\cite{Hafstad:1938,Ikeda:1968,Freer:2018,
Wei2024ClusteringProgressPerspectives}. The impact of $\alpha$ clustering on relativistic heavy-ion collisions has
been explored previously in terms of initial eccentricities, anisotropic flow,
and event-by-event fluctuations and correlations in light nuclear
systems~\cite{Broniowski2014AlphaClustering,Zhang2017ClusterFlow,
Li2020O16AlphaClustering,Ma2020AnisotropyAlphaClusteredC12Au197,
Ma2024AlphaClusteringReview}. From a first-principles perspective, these
clustering correlations are successfully described by modern \textit{ab
initio} frameworks, including Quantum Monte Carlo
methods~\cite{Carlson:2015RMP}, the No-Core Shell
Model~\cite{Barrett:2013PPNP}, Lattice Effective Field
Theory~\cite{Lee:2009PPNP,Epelbaum:2011PRL}, and advanced generator
coordinate methods~\cite{Yao:2020PRL}. Dedicated runs of
$^{16}$O+$^{16}$O and $^{20}$Ne+$^{20}$Ne collisions at
$\sqrt{s_{\mathrm{NN}}}=5.36$~TeV at the LHC therefore provide a unique
opportunity to probe $\alpha$ clustering in a new, high-temperature
regime~\cite{Brewer:2021kiv,LHC:2025LightIon}.

G. Giacalone and E. Speranza~\cite{Giacalone:2025} have recently proposed a
new paradigm relating initial-state spin fluctuations in the colliding nuclei
to final-state spin correlations of emitted hadrons via relativistic spin
hydrodynamics~\cite{Florkowski:2019RelSpinHydro,
Bhadury:2021NewDevSpinHydro,Speranza:2021SpinHydroReview,
Huang:2025IntroSpinHydro}, introducing a frame-independent $\Lambda$-pair
angular correlation observable $v_{\Lambda}^2$ as a direct experimental
handle. Within this framework, the variance of the event-by-event net
polarization of the fireball is approximately preserved during the
hydrodynamic evolution and can be accessed experimentally through measurements
of $v_{\Lambda}^2$. This opens up the possibility of probing ground-state
nuclear spin structure in high-energy collisions. However, to date, this idea
has not been applied to investigate special nuclear ground-state spin
structures such as those arising from $\alpha$ clustering.

Motivated by these developments, in the present work we investigate how
$\alpha$-cluster-induced spin structures in $^{16}$O and $^{20}$Ne are
imprinted on the initial spin fluctuations in ultra-relativistic
nucleus-nucleus collisions. A characteristic feature of $\alpha$ clustering
is that within each $2p$-$2n$ cluster, spin and isospin combine to form an
approximate spin-isospin singlet, leading to strong local cancellation of
nucleon spins and, consequently, to short-range spin--spin anti-correlations
in the nuclear many-body wave function~\cite{Wigner:1937,Sogo:2009zv,
Tohsaki:2001PRL}. We show that these microscopic correlations lead to a
suppression of the standard deviation $\sqrt{\langle \mathcal{P}^2\rangle}$
of the initial net polarization (event by event) compared to a spherical
Woods--Saxon reference without additional intrinsic spin correlations beyond
the global \(J=0\) constraint. By introducing a suitably scaled polarization
fluctuation that removes the dominant trivial finite-size effect associated
with the global $J=0$ constraint of each nucleus, we demonstrate that
cluster-induced spin structures produce characteristic dependencies on
collision centrality and system size that can serve as quantitative
diagnostics of $\alpha$ clustering when experimentally accessible final-state
spin-correlation observables are compared with the corresponding model
predictions. In particular, we propose the ratio of the scaled polarization
fluctuation between $^{20}$Ne+$^{20}$Ne and $^{16}$O+$^{16}$O collisions as
a normalized discriminator for distinguishing different $\alpha$-cluster
configurations and for confronting \textit{ab initio} nuclear structure
calculations.

The paper is organized as follows. In Sec.~\ref{sec:method} we summarize the
theoretical framework that relates initial spin fluctuations to final-state
$\Lambda$-pair observables, describe the nuclear-structure inputs for
$^{16}\mathrm{O}$ and $^{20}\mathrm{Ne}$, and define the relevant
observables. Section~\ref{sec:results} presents our numerical results for
$^{16}\mathrm{O}+{}^{16}\mathrm{O}$ and
$^{20}\mathrm{Ne}+{}^{20}\mathrm{Ne}$ collisions at
$\sqrt{s_{\mathrm{NN}}}=5.36$~TeV. In Sec.~\ref{sec:summary} we summarize
our findings and discuss their implications for future theoretical and
experimental studies.

\section{Method}
\label{sec:method}

In this section, we outline the framework for probing initial spin 
fluctuations. We first introduce the theoretical framework connecting 
initial spin fluctuations to final-state spin polarization observables.
Then, we describe the initial nuclear structure inputs for 
$^{16}\mathrm{O}$ and $^{20}\mathrm{Ne}$, starting from \textit{ab initio} 
lattice effective field theory (NLEFT) configurations, followed by 
$\alpha$-cluster models, and a 3pf Woods--Saxon distribution.
Finally, we define the key quantities: the standard deviation 
$\sqrt{\langle \mathcal{P}^2 \rangle}$ of the initial spin polarization, the 
scaled standard deviation 
$\sqrt{\langle \mathcal{P}^2 \rangle}_{\text{scaled}}$, and the ratio of scaled   
standard deviations between different collision systems.

\subsection{Initial spin fluctuations and final-state observables}
\label{subsec:spin_model}

In this subsection, the framework of Ref.~\cite{Giacalone:2025}, which
connects initial spin fluctuations to measurable final-state
observables, is summarized and adapted to the present study.
The construction proceeds in three steps:
(i) a stochastic model for the initial spin density and the associated
event polarization,
(ii) the definition of a frame-independent observable based on
$\Lambda$-pair spin correlations, and
(iii) the discussion of spin hydrodynamics, which provides the
space--time evolution.

\textbf{Initial spin density and event polarization.}
The starting point is an initial-condition model that provides, event
by event, the transverse distribution of participant nucleons and the
corresponding spin density at midrapidity.
Following Ref.~\cite{Giacalone:2025}, the initial spin density is
encoded in a scalar field $S(\tau,{\bf x})$ and a unit vector $n^i$
specifying the orientation of the polarization,
\begin{equation}
  S^i(\tau,{\bf x}) \equiv n^i S(\tau,{\bf x}) \quad [\hbar/{\rm fm}^3] ,
  \label{eq:S_vec}
\end{equation}
and an initial spin density per unit rapidity at midrapidity is defined
as
\begin{equation}
  \mathcal{S}({\bf x}) \equiv \lim_{\tau\to 0^+} \tau S(\tau,{\bf x}) \quad [\hbar/{\rm fm}^2] .
  \label{eq:S_rapidity}
\end{equation}
In the present analysis the transverse distribution of participants is
taken from the \trento{} model~\cite{Moreland:2014oya,Moreland:2018gsh,Soeder:2023vdn}, and $\mathcal{S}({\bf x})$ is implemented in a
Glauber-type form \cite{Miller:2007ri,Alver:2008aq} as
\begin{equation}
    \mathcal{S}({\bf x}) = S_0 \frac{\hbar}{2}
    \sum_{i=1}^{N_{\text{part}}} s_i\,
    w_s({\bf x}-{\bf x}_i) ,
    \label{eq:S_field}
\end{equation}
where ${\bf x}$ is the transverse coordinate, ${\bf x}_i$ denotes the
position of the $i$-th participant, $w_s$ is a normalized smearing
profile, $s_i=\pm1$ is a spin projection along a quantization axis, and
$S_0$ is a dimensionless normalization parameter.
The orientation of the quantization axis is chosen randomly and
independently in each event, such that the ensemble of events is
globally unpolarized.

Integrating $\mathcal{S}({\bf x})$ over the transverse plane gives the
net spin carried by the participants in that event.
It is convenient to characterize the corresponding fluctuating
polarization through the dimensionless event variable
\begin{equation}
    \mathcal{P}_{\text{ini}} =
    \frac{1}{N_{\text{part}}}
    \sum_{i=1}^{N_{\text{part}}} s_i ,
    \qquad -1\le\mathcal{P}_{\text{ini}}\le 1 .
    \label{eq:Pini_def}
\end{equation}
For colliding nuclei with total angular momentum $J=0$, and with the
quantization axis randomized from event to event, the mean polarization
vanishes, $\langle\mathcal{P}_{\text{ini}}\rangle=0$, whereas the
variance $\langle\mathcal{P}_{\text{ini}}^2\rangle$ is nonzero and
measures the size of initial spin fluctuations.
The field $\mathcal{S}({\bf x})$ therefore provides a well-defined
initial spin density profile that can serve as input for spin
hydrodynamics, while $\mathcal{P}_{\text{ini}}$ encodes the associated
event-wise net polarization.

\textbf{\texorpdfstring{$\Lambda$}{Lambda}-pair spin correlations.}
The direction of $\mathcal{P}_{\rm ini}$ is random from event to event.
Consequently, the global polarization with respect to a fixed laboratory axis
vanishes after event averaging, and observables linear in $\mathcal{P}$
are blind to the fluctuations of interest.
Unlike conventional global-polarization measurements that probe the mean
$\langle\mathcal{P}\rangle$ (often discussed in connection with thermal
vorticity and shear), we focus on event-by-event fluctuations quantified
by $\langle\mathcal{P}^2\rangle$, accessible through two-particle spin
correlations.
One therefore needs an observable sensitive to $\mathcal{P}^2$ rather
than $\mathcal{P}$.

Ref.~\cite{Giacalone:2025} proposed to access initial spin fluctuations
via spin correlations of pairs of $\Lambda$ hyperons reconstructed in
the final state. Owing to the parity-violating weak decay
$\Lambda\to p\pi^-$, the distribution of the decay proton direction
$\hat{\mathbf{p}}_p$ in the $\Lambda$ rest frame is correlated with the
$\Lambda$ polarization vector $\mathbf{P}$ as
\begin{equation}
  \frac{dN}{d\Omega_p}\propto
  1+\alpha_\Lambda\,\mathbf{P}\cdot\hat{\mathbf{p}}_p ,
  \label{eq:Lambda_decay_main}
\end{equation}
where $\alpha_\Lambda$ is the weak-decay parameter~\cite{ParticleDataGroup:2024cfk}
and $d\Omega_p$ is the solid angle of $\hat{\mathbf{p}}_p$. An analogous relation holds for anti-hyperons
$\bar\Lambda\to\bar p\pi^+$ with decay parameter
$\alpha_{\bar\Lambda}\simeq -\alpha_\Lambda$.
This relation forms the basis of hyperon-polarization measurements
in heavy-ion collisions~\cite{Adamczyk:2017}.

In a given event we consider pairs of (anti)hyperons, labeled 1 and~2,
which can be of the type $\Lambda\Lambda$, $\Lambda\bar\Lambda$, or
$\bar\Lambda\bar\Lambda$.
Their decay (anti)protons define two unit vectors
$\hat{\mathbf{p}}_{p,1}$ and $\hat{\mathbf{p}}_{p,2}$ in the respective
rest frames, and we denote by
\begin{equation}
  \cos\Delta\theta
  \equiv \hat{\mathbf{p}}_{p,1}\cdot\hat{\mathbf{p}}_{p,2}
  \label{eq:Delta_theta_main}
\end{equation}
the cosine of the relative angle between them.
By averaging $\cos\Delta\theta$ over all hyperon pairs and events in a
given centrality class and normalizing by the appropriate weak-decay
parameters $\alpha_1$ and $\alpha_2$ of the two hyperons in the pair,
one defines the frame-independent correlator
\begin{equation}
    v_\Lambda^2 \equiv
    \frac{9}{\alpha_1\alpha_2}
    \Big\langle\!\Big\langle
      \cos\Delta\theta
    \Big\rangle\!\Big\rangle ,
    \label{eq:vLambda_def_main}
\end{equation}
where the double brackets denote a pair and event average.
In practice, $\alpha_1$ and $\alpha_2$ are taken as the corresponding
decay parameters of each hyperon in the pair, so that the same
definition applies uniformly to $\Lambda\Lambda$, $\Lambda\bar\Lambda$,
and $\bar\Lambda\bar\Lambda$ pairs.

This construction is directly analogous to methods employed more broadly
in high-energy physics, where angular correlations of decay products are
used to access underlying spin correlations and, in suitable settings,
quantum entanglement, both in hyperon systems and in top--antitop
production at the LHC~\cite{Sheng:2025puj,Baumgart:2012ay,ATLAS:2023fsd}.

Under the assumptions of Ref.~\cite{Giacalone:2025}---namely that in
each event all hyperons share a common polarization vector
$\mathbf{P}$, whose orientation is random from event to event, and that
nonflow correlations between different hyperons are negligible---the
observable $v_\Lambda^2$ reduces to the variance of the event-wise net
polarization in the final state,
\begin{equation}
    v_\Lambda^2 = \langle\mathcal{P}^2\rangle_{\text{final},\Lambda} .
    \label{eq:vLambda_P2_main}
\end{equation}
The subscript ``$\Lambda$'' indicates that the correlator is constructed
from reconstructed (anti)$\Lambda$ hyperons in the final state.
The standard derivation of Eq.~\eqref{eq:vLambda_P2_main}, starting from
the joint angular distribution of the two decay protons and the
spin-correlation matrix, is given in
Ref.~\cite{Giacalone:2025} and in the Supplemental Material therein.
In our framework $v_\Lambda^2$ provides an experimental handle on the
initial polarization variance
$\langle\mathcal{P}_{\text{ini}}^2\rangle$ generated by different
nuclear spin structures, under the dynamical assumptions discussed
below.

\textbf{Spin hydrodynamics and connection to initial fluctuations.}
The space--time evolution of the quark--gluon plasma with spin degrees
of freedom is described by relativistic spin hydrodynamics~\cite{Florkowski:2019RelSpinHydro,Speranza:2021SpinHydroReview,becattiniPolarizationVorticityQuark2020}.
In the working pseudo-gauge adopted here, where $T^{\mu\nu}$ is taken
to be symmetric, the equations of motion for an uncharged fluid are
given by the conservation of energy--momentum and of the spin tensor
$S^{\lambda,\mu\nu}$,
\begin{equation}
    \partial_\mu T^{\mu\nu}=0,\qquad
    \partial_\lambda S^{\lambda,\mu\nu}=0 ,
    \label{eq:hydro_EOM_main}
\end{equation}
where $T^{\mu\nu}$ is the energy--momentum tensor and
$S^{\lambda,\mu\nu}$ is antisymmetric in the last two indices.
Different pseudo-gauge choices correspond to different decompositions
of orbital and spin angular momentum, while leaving the total conserved
charges unchanged~\cite{Weickgenannt:2022PseudoGauge,Hongo:2021SpinHydroTorsion}.
More generally, total angular-momentum conservation implies
$\partial_\lambda S^{\lambda,\mu\nu}=T^{\nu\mu}-T^{\mu\nu}$, so that
Eq.~(\ref{eq:hydro_EOM_main}) corresponds to pseudo-gauges in which
$T^{\mu\nu}$ is symmetric; see
Appendix~\ref{app:spin_hydro}, Eq.~\eqref{eq:spin_noncons_app}.

Following Ref.~\cite{Giacalone:2025}, one may define from
$S^{\lambda,\mu\nu}$ a spin-density four-vector $S^\mu$ and
operationally relate its spatial components to the scalar field
$S({\bf x})$ introduced above.
Here $S^{\lambda,\mu\nu}$ denotes the spin tensor,
$S^\mu$ the associated spin-density four-vector,
$S(\tau,{\bf x})$ the scalar amplitude entering its spatial components,
and $\mathcal{S}({\bf x})$ the corresponding initial density per unit
rapidity at midrapidity.
In the present work, this relation is used only as an operational
mapping between the initial-state model and the spin-hydrodynamic
description.
The corresponding conventions and working assumptions, including the
role of pseudo-gauge choices and the mapping between
$S^{\lambda,\mu\nu}$ and the initial density $\mathcal{S}({\bf x})$,
are summarized in Appendix~\ref{app:spin_hydro}.

In the regime relevant for ultra-relativistic nucleus--nucleus
collisions, Ref.~\cite{Giacalone:2025} assumes that the total spin of
the fireball is approximately conserved during the hydrodynamic
evolution and that spin diffusion is moderate.
Under this assumption the magnitude of the event polarization is
essentially preserved from the initial to the final state, so that here
$\langle\mathcal{P}^2\rangle_{\text{final}}$ denotes the net
polarization variance of the overall evolving fireball (before
restricting to a specific hadron species).
\begin{equation}
    \langle\mathcal{P}^2\rangle_{\text{final}}
    \simeq
    \langle\mathcal{P}_{\text{ini}}^2\rangle ,
    \label{eq:P2_preserve}
\end{equation}
While the relation above concerns the net polarization variance of the
evolving fireball, the experimentally accessible quantity is
reconstructed from a subset of final-state hadrons, here
(anti)$\Lambda$ hyperons.
We denote the measured correlator by
$v_\Lambda^2 \equiv \langle\mathcal{P}^2\rangle_{\text{final},\Lambda}$.
This should be distinguished from
$\langle\mathcal{P}^2\rangle_{\text{final}}$, since the final-state spin
content is distributed among different hadron species and the
reconstructed $\Lambda$ sample is affected by hadronization and decays
(including resonance feed-down)~\cite{Becattini:2019FeedDownTransfer,Xia:2019,Li:2021_MultiStrangeFeeddown,Becattini:2017}.
Schematically, one may write
$\langle\mathcal{P}^2\rangle_{\text{final},\Lambda} \sim
k_\Lambda^2\,\langle\mathcal{P}^2\rangle_{\text{ini}}+\Delta^{(\Lambda)}$,
where $k_\Lambda$ is an effective, species-dependent
transfer/attenuation factor, and $\Delta^{(\Lambda)}$ represents
additional contributions beyond the initial-state fluctuations.

Beyond the initial spin fluctuations, the measured two-particle
correlator may also receive other contributions.
Long-range contributions can arise from velocity-gradient induced
effects (thermal vorticity and shear); at LHC energies and in more
central collisions (with smaller global angular momentum), such
contributions are expected to be reduced (parametrically suppressed) by
the small $\omega/T$ scale~\cite{Giacalone:2025,Speranza:2021SpinHydroReview,becattiniPolarizationVorticityQuark2020}.
Short-range background correlations, including nonflow-like effects,
can also affect the two-particle correlator.
Additional short-range dynamical sources may originate from
short-distance fluctuations of strong-interaction fields, often modeled
as vector-meson ($\phi$) fields~\cite{Wang:2023fvy,Sheng:2025puj};
being localized, such contributions can be diagnosed and typically
reduced by standard pair-separation systematics in two-particle
measurements, e.g.\ by imposing a pseudorapidity-gap selection and/or
checking the dependence on the pair separation.

A more quantitative assessment and systematic separation of the various
contributions to the measured spin correlator will ultimately require
full event-by-event simulations of relativistic spin hydrodynamics
or spin transport, which we leave for future work.
In the present study, we instead concentrate on the initial
polarization fluctuations generated by different microscopic nuclear
structures, encoded in $\langle\mathcal{P}_{\text{ini}}^2\rangle$ (and
its scaled version), and use ratios between collision systems as a
strategy to reduce sensitivity to residual dynamical effects.
Under the working assumption that such residual effects are largely
common-mode between $^{16}$O and $^{20}$Ne under the same collision
energy and analysis cuts, the system-size ratio introduced below is
expected to mitigate a significant part of these effects, thereby
providing a cleaner sensitivity to the underlying nuclear spin
structure.

\subsection{Nuclear structure inputs}
\label{subsec:nuclear_structure}

To investigate how nuclear structure affects initial spin fluctuations,
we focus on the spin structure of the ground states of
${}^{16}\mathrm{O}$ and ${}^{20}\mathrm{Ne}$.
To construct initial conditions for spin hydrodynamics, the model must
specify, on an event-by-event basis, both the spatial coordinates and
the spin of all nucleons.
${}^{16}\mathrm{O}$ and ${}^{20}\mathrm{Ne}$ belong to the class of
self-conjugate $4n$ light
nuclei, for which mean-field descriptions are known to be strongly
modified by the formation of spatially localized $\alpha$ clusters
\cite{Hafstad:1938,Ikeda:1968,Freer:2018}.
A characteristic feature of such clustering is that, inside each
$\alpha$ particle, the spins and isospins of the four nucleons combine
into (approximately) spin--isospin singlet configurations~\cite{Wigner:1937,Sogo:2009zv,Tohsaki:2001PRL}, leading to
strong local cancellation of spin and, consequently, pronounced
short-range spin--spin anti-correlations in the nuclear many-body
wave function.
To explore how these correlations are imprinted on the
initial spin polarization, the model assigns to each nucleon a spatial
position $\mathbf{x}_i$ and a spin projection $s_i=\pm1$, and employs
three distinct types of nuclear-structure input: \textit{ab initio}
configurations, explicit $\alpha$-cluster models, and a baseline
3pf Woods--Saxon distribution without clustering.

\textbf{Ab initio NLEFT configurations.}
For ${}^{16}\mathrm{O}$ and ${}^{20}\mathrm{Ne}$ we employ microscopic
configurations generated within Nuclear Lattice Effective Field Theory
(NLEFT)~\cite{Lee:2009PPNP,Lahde:2019NLEFTreview}.~NLEFT solves the nuclear $A$-body problem on a spacetime lattice using
chiral effective field theory interactions, and, in particular, the
pinhole algorithm allows one to sample the $A$-body density in both
coordinate and spin--isospin space~\cite{Elhatisari:2017eno,Giacalone:2024luz,Summerfield:2021oex}.~In this work we use the ${}^{16}\mathrm{O}$ and ${}^{20}\mathrm{Ne}$
pinhole configurations reported in Refs.~\cite{Giacalone:2024luz,Summerfield:2021oex}.~These configurations provide, for each Monte-Carlo event, the full set
of nucleon positions together with their spin--isospin quantum numbers,
so that correlations between positions and spins, including higher-order
many-body correlations, are preserved and can be directly mapped to the initial spin-density profiles used in our model.
Note that the NLEFT pinhole output stores spin quantum numbers as projections
along a fixed quantization axis (an $s_z$-type projection). In our implementation
each nucleus is randomly rotated event by event, so that this quantization
direction is sampled isotropically in the laboratory frame; we therefore use
these two-valued projections as the spin labels entering scalar observables
such as $s_i s_j$ and $\mathcal{P}$ (see Appendix~\ref{app:NLEFT} for details
and limitations).

\begin{figure}[t]
\begin{center}
\includegraphics[width=0.95\linewidth]{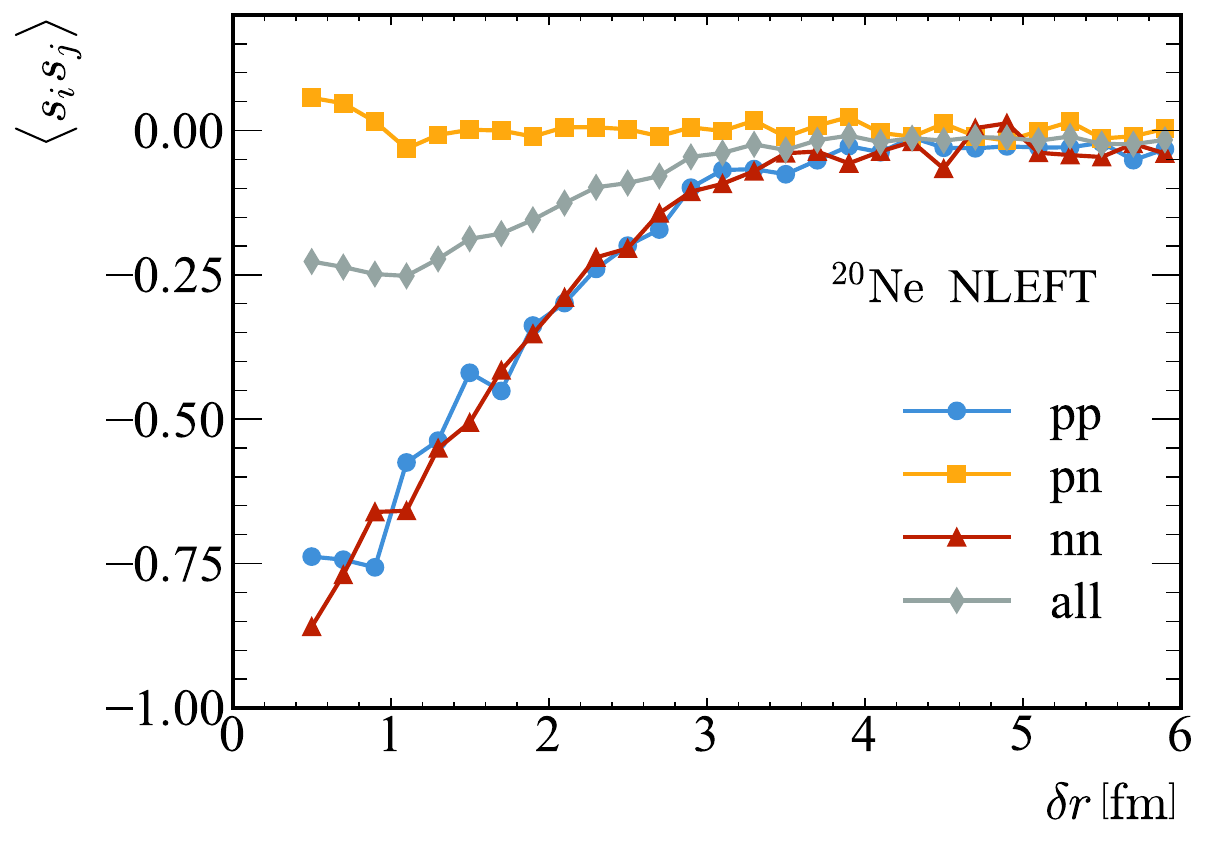}
\end{center}
\caption{(Color online) Two-nucleon spin correlation $\langle s_i s_j \rangle$ as a function of the relative distance $\delta r$ in $^{20}\mathrm{Ne}$ from \textit{ab initio} NLEFT calculations. 
The blue circles, yellow squares, red triangles, and gray diamonds represent $\langle s_i s_j \rangle(\delta r)$ for proton--proton, proton--neutron, neutron--neutron, and all nucleon pairs, respectively.}
\label{fig:sisj}
\end{figure}

The microscopic spin structure captured by NLEFT is illustrated in Fig.~\ref{fig:sisj}, which shows the two-nucleon spin correlation 
$\langle s_i s_j \rangle$ as a function of relative distance $\delta r$ 
in $^{20}\mathrm{Ne}$. 
In practice, $\langle s_i s_j \rangle(\delta r)$ is obtained by taking, for each NLEFT configuration,
the product $s_i s_j$ of all nucleon pairs, binning them according to their separation
$\delta r = |{\bf x}_i-{\bf x}_j|$, and then averaging over the ensemble of configurations. 
A distinct pattern is observed: for like-particle 
pairs (proton--proton and neutron--neutron), the spin correlation is 
negative at short distances ($\delta r \lesssim 3$ fm) and vanishes at 
larger distances. This reflects the Pauli exclusion principle and the 
tendency of identical nucleons to pair with opposite spins within 
compact clusters. Conversely, the proton--neutron correlation remains 
negligible across all distances. This specific short-range 
spin anti-correlation is consistent with the expected 
$\alpha$-cluster structure of $^{20}\mathrm{Ne}$, 
which strongly deviates from uncorrelated mean-field densities,
see, e.g., Refs.~\cite{Yamaguchi:2023,Bijker:2021Ne20}.

\textbf{Explicit $\alpha$-cluster models.} 

\begin{figure*}[!tbp]
\begin{center}
\includegraphics[width=0.95\linewidth]{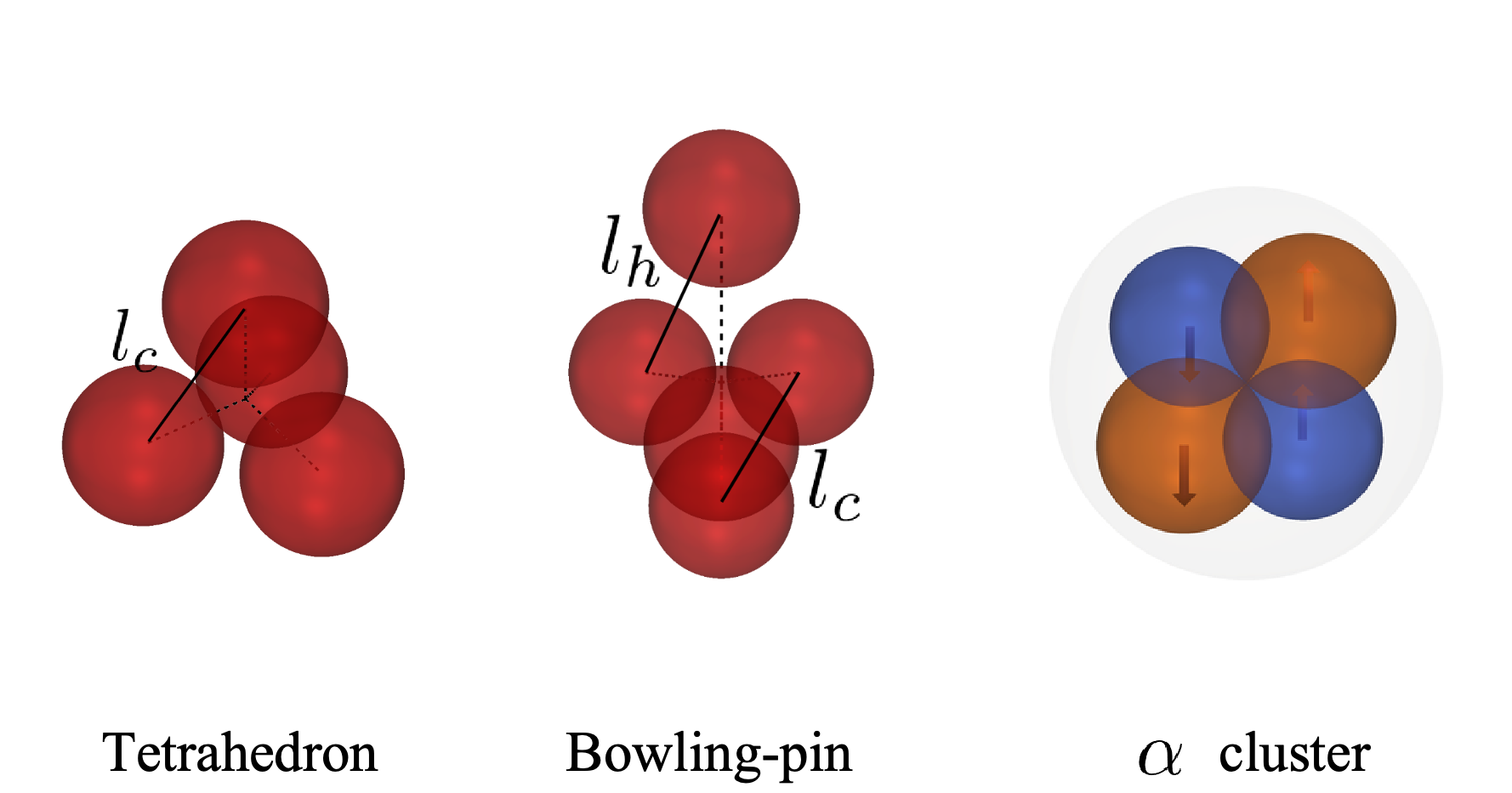}
\end{center}
\caption{
Schematic illustration of the $\alpha$-cluster geometries 
implemented in the model.
(Left) The Tetrahedron configuration of $^{16}\mathrm{O}$ is 
characterized by the $\alpha$-cluster edge length $\ell_c$.
(Middle) The Bowling-pin configuration of $^{20}\mathrm{Ne}$ is 
characterized by the core parameter $\ell_c$ and the distance $\ell_h$ 
between the fifth cluster and the core.
(Right) A schematic of the $\alpha$ cluster, which contains 
two protons (orange spheres) and two neutrons (blue spheres). 
The directions of the black arrows ($\uparrow\downarrow$)  in the right panel indicate
the spin orientations of the individual nucleons.
The spins of the two protons cancel each other out, as do the 
spins of the two neutrons, resulting in a net-zero spin for the cluster.
Each red sphere in the geometry plots (Left and Middle) 
represents an $\alpha$ cluster.
}

\label{fig:geometry}
\end{figure*}

To systematically study the influence of geometry and spin suppression, 
we adopt phenomenological $\alpha$-cluster models that have been shown to 
reproduce one-body densities of $^{16}\mathrm{O}$ and $^{20}\mathrm{Ne}$ obtained from modern 
\textit{ab initio} calculations. 
Following the Gaussian cluster framework of Ref.~\cite{Mehrabpour:2025alpha}, 
each nucleus is composed of $N_\alpha$ independent $\alpha$ particles. 
The spatial distribution of nucleons within each cluster $i$ is sampled from a Gaussian profile,
\begin{equation}
    \rho_{\alpha}(\mathbf{x}) \propto 
    \exp\Big[-\frac{3(\mathbf{x}-\mathbf{L}_i)^2}{2r_L^2}\Big],
\end{equation}
where $\mathbf{L}_i$ is the center of the $i$-th cluster and $r_L$ is 
the root mean-square radius of cluster (RMS). 
To model the spin--isospin singlet structure of an $\alpha$ cluster, we enforce strict spin--isospin cancellation within each cluster:
each $\alpha$ cluster consists of two protons and two neutrons, with spins strictly anti-aligned 
($\uparrow\downarrow$) for each isospin pair, resulting in zero net spin per cluster.

For the geometry, we consider structures that are consistent with both low-energy $\alpha$-cluster 
models~\cite{Yamaguchi:2023} and recent applications to relativistic collisions~\cite{Li:2025NeNe,Mehrabpour:2025alpha}, 
as illustrated in Fig.~\ref{fig:geometry}:
\begin{itemize}
    \item \textbf{$^{16}\mathrm{O}$ (tetrahedron):} Four $\alpha$ clusters 
    are arranged at the vertices of a regular tetrahedron with side length 
    $\ell_c$.
    \item \textbf{$^{20}\mathrm{Ne}$ (bowling--pin):} Modeled as a 
    $^{16}\mathrm{O}$ tetrahedral core plus an additional $\alpha$ cluster. 
    This forms a distorted bi-pyramidal or ``bowling--pin'' shape, 
    characterized by the core size $\ell_c$ and the distance $\ell_h$ 
    of the fifth cluster from the core center.
\end{itemize}

In practice, we do not refit the cluster parameters in this work. 
Instead, we take over the parameter sets $(r_L,\ell_c,\ell_h)$ obtained in 
Ref.~\cite{Mehrabpour:2025alpha}, where they were determined by minimizing 
a $\chi^2$ function to reproduce the one-body nucleon densities of 
NLEFT, VMC, PGCM and of a three-parameter Fermi (3pf) reference density 
for $^{16}\mathrm{O}$ and $^{20}\mathrm{Ne}$. 
The resulting values for the configurations used in this study are collected 
in TABLE~\ref{tab:cluster_params}. 

In addition to these pure $\alpha$-cluster geometries, we also consider a 
schematic $^{16}\mathrm{O}+\alpha$ configuration for $^{20}\mathrm{Ne}$, 
motivated by the $^{16}\mathrm{O}+\alpha$  structures discussed in 
Refs.~\cite{Yamaguchi:2023,Li:2025NeNe}. 
In this case, a spherical $^{16}\mathrm{O}$ core is generated from a 3pf 
Woods--Saxon distribution and a single $\alpha$ cluster with RMS $r_L$ is attached at a 
fixed distance $d_1$ along a random direction in space. 
The parameters of the $\alpha$-cluster models, of the 3pf Woods--Saxon nuclei, 
and of the $^{16}\mathrm{O}+\alpha$ configuration are collected in 
TABLE~\ref{tab:cluster_params}.

\begin{table}[t]
\centering
\caption{Nuclear-structure parameters used in this work. 
The upper part lists the $\alpha$-cluster model parameters 
$(r_L, \ell_c, \ell_h)$ in fm for $^{16}\mathrm{O}$ and 
$^{20}\mathrm{Ne}$, adopted from the $\chi^2$ fits of 
Ref.~\cite{Mehrabpour:2025alpha} to NLEFT, PGCM, VMC and 3pf densities. 
The lower part summarizes the parameters of the three-parameter Fermi
(3pf) Woods--Saxon distributions and of the schematic $^{16}\mathrm{O}+\alpha$ configuration.}
\label{tab:cluster_params}

\begin{tabular*}{\linewidth}{@{\extracolsep{\fill}}lcccc@{\extracolsep{\fill}}}
\hline \hline
Nucleus / Model & Reference Density & $r_L$ (fm) & $\ell_c$ (fm) & $\ell_h$ (fm) \\ \hline
$^{16}\mathrm{O}$ (tetrahedron) & NLEFT & 1.84 & 3.17 & -- \\
                                & PGCM  & 1.88 & 3.06 & -- \\
                                & VMC   & 1.52 & 3.26 & -- \\ \hline
$^{20}\mathrm{Ne}$ (bowling--pin) & NLEFT & 2.20 & 3.00 & 3.50 \\
\hline \hline
\end{tabular*}

\vspace{0.25cm}

\begin{tabular*}{\linewidth}{@{\extracolsep{\fill}}lccccc@{\extracolsep{\fill}}}
\hline \hline
Nucleus / Model & $R$ (fm) & $a$ (fm) & $w$ & $d_1$ (fm) & $r_L$ (fm) \\ \hline
$^{16}\mathrm{O}$ (3pf)                 & 2.608 & 0.513 & $-0.051$ & --   & --    \\
$^{20}\mathrm{Ne}$ (3pf)                & 2.791 & 0.698 & $-0.168$ & --   & --    \\
$^{20}\mathrm{Ne}$ ($^{16}\mathrm{O}+\alpha$) 
                                        & 2.608 & 0.513 & $-0.051$ & 3.0  & 2.164 \\
\hline \hline
\end{tabular*}

\end{table}

\textbf{3pf Woods--Saxon reference.}
As a reference for comparison, we also employ spherical three-parameter Fermi
(3pf) Woods--Saxon charge-density distributions.
The radial density is taken as
\begin{equation}
  \rho(r)
  = \rho_0\,
    \frac{1 + w \left(\dfrac{r}{R}\right)^2}
         {1 + \exp\!\left(\dfrac{r - R}{a}\right)} ,
\end{equation}
where $R$ is the half-density radius, $a$ is the surface diffuseness,
$w$ controls a possible central depression, and $\rho_0$ is fixed by the
normalization to the mass number $A$.
For $^{16}\mathrm{O}$ and $^{20}\mathrm{Ne}$ we use the standard 3pf parameter sets
extracted from elastic electron-scattering data in Ref.~\cite{DeVries:1987-3pf}.
In this model, nucleon positions are sampled from the $\rho(r)$ without any geometric clustering. Spins are assigned by
enforcing the global $J=0$ constraint in each nucleus: we sample exactly
$A_{\text{mass}}/2$ spins with $+1$ and $A_{\text{mass}}/2$ spins with $-1$
and then randomly distribute them among the nucleons.
This reference configuration contains no spin--position correlations, apart from the
same-nucleus anticorrelation induced by conditioning on the global $J=0$ constraint.

\subsection{Spin fluctuations, baseline, and scaled observables}
\label{subsec:observables}

To quantify the imprint of nuclear structure on spin observables,
we relate initial polarization fluctuations to the microscopic two-body
spin densities of the colliding nuclei. A detailed derivation is provided
in Appendix~\ref{app:P2_spin}.

\textbf{Net polarization and spin densities.}
The event-by-event net polarization $\mathcal{P}$ is defined as the
average spin of all participants,
\begin{equation}
    \mathcal{P}
    = \frac{1}{N_{\text{part}}}
      \left( \sum_{i=1}^{N_{\text{part}}^A} s_{i,A} +
             \sum_{j=1}^{N_{\text{part}}^B} s_{j,B} \right),
    \label{eq:P_def}
\end{equation}
where $s_{i,A}=\pm 1$ and $s_{j,B}=\pm 1$ denote the spin projections of
participants from nuclei $A$ and $B$, and
$N_{\text{part}} = N_{\text{part}}^A + N_{\text{part}}^B$.
For unpolarized colliding ions one has
$\langle s_{i,A}\rangle = \langle s_{j,B}\rangle = 0$, hence
$\langle \mathcal{P} \rangle = 0$.~In the present setup this participant-level quantity is identical to $\mathcal{P}_{\text{ini}}$ defined in Eq.~\eqref{eq:Pini_def},
and we use the two notations interchangeably depending on context.

Assuming that spin degrees of freedom in the two nuclei are independent,
and for the sake of a schematic discussion neglecting event-by-event
participant-number fluctuations within a given centrality class,
the variance $\langle \mathcal{P}^2 \rangle$ can be expressed in terms of
one- and two-body spin densities in each nucleus. For a system with
$N$ participants this dependence can be written schematically as
\begin{equation}
    \langle \mathcal{P}^2 \rangle \propto
    N \sum_s \rho^{(1)}(s)\, s^2
    + N(N/2-1) \sum_{s_1,s_2} \rho^{(2)}(s_1,s_2)\,s_1 s_2 ,
    \label{eq:P2_rho_main}
\end{equation}
where $\rho^{(1)}$ and $\rho^{(2)}$ denote, respectively, the effective one-
and two-body spin densities of the participants (as defined in
Appendix~\ref{app:P2_spin}).
The first term encodes the trivial variance of independent spins,
while the second term is sensitive to genuine two-body spin--spin
correlations, such as those induced by $\alpha$ clustering.
In practice, our numerical evaluation below retains the event-by-event
$N_{\text{part}}^{A,B}$ fluctuations through the \trento{} output and the
adopted centrality definition.

\textbf{Null baseline: global $J=0$ constraint and no additional spin correlations.}
We construct a null baseline using the event-by-event $N_{\text{part}}^{A,B}$
from the \trento{} model together with a minimal spin-correlation prescription.
Nuclei such as $^{16}$O and $^{20}$Ne have a $J^{\pi}=0^{+}$ assignment.
In our simplified representation with $s_i=\pm 1$, we implement this by imposing
the global constraint $\sum_{i=1}^{A_{\text{mass}}} s_i = 0$, i.e., random spin
assignments conditioned on $J=0$ and without additional intrinsic spin correlations.

For a nucleus with mass number $A_{\text{mass}}$ one finds (see
Appendix~\ref{app:P2_spin})
\begin{equation}
c= \langle s_i s_j \rangle_{\text{base}}
= -\frac{1}{A_{\text{mass}} - 1}, \qquad i\neq j ,
\label{eq:base_pair}
\end{equation}
which is the uniform negative background correlation induced by conditioning
on the global $J=0$ constraint within this permutation-symmetric random-assignment
ensemble.

Assuming that this background correlation characterizes all same-nucleus pairs,
the corresponding baseline variance of the polarization in a collision between
nuclei $A$ and $B$ is
\begin{equation}
\langle\mathcal{P}^2\rangle_{\text{base}}
=
\Bigg\langle
\frac{1}{\Npart^{2}}
\Big[
  \Npart
  + c \sum_{X=A,B}
  N_{\text{part}}^{X}\big(N_{\text{part}}^{X}-1\big)
\Big]
\Bigg\rangle .
\label{eq:P2_base_main}
\end{equation}
By construction, $\langle \mathcal{P}^2 \rangle_{\text{base}}$ is a theory-defined reference that implements the global $J=0$ constraint and retains the contribution from event-by-event $N_{\text{part}}^{A,B}$ fluctuations, while containing no additional intrinsic spin correlations.
For each nuclear-structure input considered in this work, $\langle\mathcal{P}^2\rangle_{\rm base}$ and the scaled quantity defined below
are evaluated on the same event sample and with the same centrality definition, so that the remaining model uncertainty is expected to be dominated by
the model estimate of $N_{\rm part}^{A,B}$ and by centrality-selection (mapping) fluctuations~\cite{Zhou:2018zxt,Loizides:2025OONeNe}.

\textbf{Scaled spin fluctuations and system-size ratio.}
Since $\langle \mathcal{P} \rangle = 0$, the standard deviation of the
event-wise polarization is simply
\begin{equation}
    \mathrm{std}(\mathcal{P})
    = \sqrt{\langle \mathcal{P}^2 \rangle} .
\end{equation}
To suppress the dominant trivial system-size/participant-number effects
associated with the global $J=0$ constraint and to highlight nontrivial spin
structures, we define the scaled spin fluctuation as
\begin{equation}
    \bigl(\sqrt{\langle \mathcal{P}^2 \rangle}\bigr)_{\text{scaled}}
    \equiv
    \frac{\sqrt{\langle \mathcal{P}^2 \rangle}}
         {\sqrt{\langle \mathcal{P}^2 \rangle_{\text{base}}}} ,
    \label{eq:P2_scaled_def}
\end{equation}
which equals unity in the null baseline of Eq.~\eqref{eq:P2_base_main} and
deviates from unity when additional spin--spin correlations are present, such as
local spin cancellation inside $\alpha$ clusters.
Throughout this work, Eq.~\eqref{eq:P2_scaled_def} is applied to the
participant-level variance $\langle\mathcal{P}^2\rangle$ computed from the
nuclear spin configurations; the relation to experimentally reconstructed
correlators (e.g.\ from (anti)$\Lambda$ hyperons) is discussed in
Sec.~\ref{subsec:spin_model}.

Finally, to compare different collision systems we introduce the central
comparison quantity used in Sec.~\ref{sec:results}, namely the ratio of scaled fluctuations
between $^{20}\mathrm{Ne}+{}^{20}\mathrm{Ne}$ and $^{16}\mathrm{O}+{}^{16}\mathrm{O}$
collisions,
\begin{equation}
R_{^{20}\mathrm{Ne}/^{16}\mathrm{O}}
\equiv
\frac{
\left(\sqrt{\langle \mathcal{P}^{2}\rangle}\right)_{\text{scaled}}^{\,
^{20}\mathrm{Ne}+{}^{20}\mathrm{Ne}}
}{
\left(\sqrt{\langle \mathcal{P}^{2}\rangle}\right)_{\text{scaled}}^{\,
^{16}\mathrm{O}+{}^{16}\mathrm{O}}
} .
\label{eq:R_ratio_def}
\end{equation}
Because $^{16}$O and $^{20}$Ne are nearby light systems, we adopt the working assumption that residual dynamical and reconstruction effects
are largely common-mode under the same collision energy and analysis cuts, and thus their impact is reduced in the ratio.
Using scaled rather than raw fluctuations additionally factors out the dominant $J=0$-conditioned finite-$N_{\text{part}}$ suppression, so that
$R_{^{20}\mathrm{Ne}/^{16}\mathrm{O}}$ is more directly sensitive to additional intrinsic spin correlations, such as cluster-induced local spin cancellation.
While model-inferred quantities (e.g.\ $N_{\text{part}}^{A,B}$ and the centrality selection/mapping) entail modeling dependence and associated uncertainties,
evaluating both the scaling and the ratio within the same framework helps reduce sensitivity to these correlated systematics.

\section{Numerical results}
\label{sec:results}

In this section, we present numerical simulations of initial spin
fluctuations in $^{16}\mathrm{O}+{}^{16}\mathrm{O}$ and
$^{20}\mathrm{Ne}+{}^{20}\mathrm{Ne}$ collisions at
$\sqrt{s_{\mathrm{NN}}}=5.36~\mathrm{TeV}$, based on the framework
outlined in Sec.~\ref{subsec:spin_model}.
For each nuclear-structure model introduced in Sec.~\ref{subsec:nuclear_structure},
we generate a large ensemble of \trento{} events, compute the event-wise
polarization $\mathcal{P}$, and study both the standard deviation
$\sqrt{\langle \mathcal{P}^{2}\rangle}$ and the scaled quantity
$\left(\sqrt{\langle \mathcal{P}^{2}\rangle}\right)_{\text{scaled}}$
defined in Sec.~\ref{subsec:observables}.
In addition, we analyze the system-size ratio of the scaled fluctuations
between $^{20}\mathrm{Ne}+{}^{20}\mathrm{Ne}$ and
$^{16}\mathrm{O}+{}^{16}\mathrm{O}$ collisions, which provides a potentially cleaner discriminator of differences in the underlying spin structure.
The results are shown as a function of centrality, which is determined
from the total entropy in \trento{}~\cite{Moreland:2014oya,Moreland:2018gsh}, in analogy with
Ref.~\cite{Giacalone:2025}.

As an unclustered nuclear-structure reference, we employ spherical 3pf Woods--Saxon distributions for both systems.
Deviations from this 3pf Woods--Saxon reference therefore quantify additional spin--spin correlations induced by cluster structures.

\subsection{\texorpdfstring{$^{20}\mathrm{Ne}+{}^{20}\mathrm{Ne}$}{Ne+Ne} collisions}

\begin{figure}[t]
\begin{center}
\includegraphics[width=0.95\linewidth]{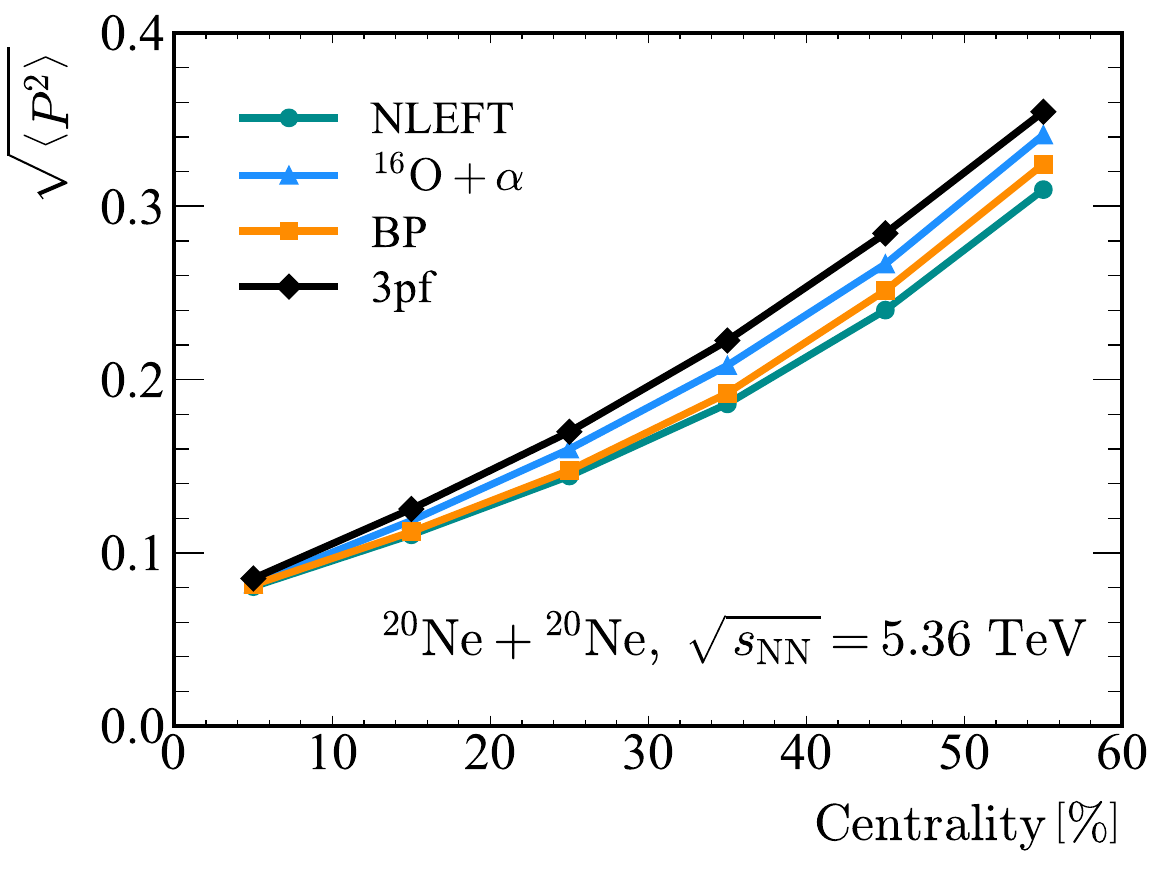}
\end{center}
\caption{(Color online) Standard deviation of the polarization parameter 
$\sqrt{\langle \mathcal{P}^{2} \rangle}$ as a function of centrality
in $^{20}\mathrm{Ne}+{}^{20}\mathrm{Ne}$ collisions at 
$\sqrt{s_{\mathrm{NN}}}=5.36~\mathrm{TeV}$ for different nuclear-structure 
configurations. The green circles, blue triangles, orange squares, and black 
diamonds correspond to the NLEFT, $^{16}\mathrm{O}+\alpha$, BP, and spherical 
3pf Woods--Saxon configurations, respectively.}
\label{fig:std_p2_Neon}
\end{figure}

Fig.~\ref{fig:std_p2_Neon} displays the centrality dependence of
$\sqrt{\langle \mathcal{P}^{2}\rangle}$ in $^{20}\mathrm{Ne}+{}^{20}\mathrm{Ne}$ collisions for four different nuclear configurations: the \textit{ab initio} NLEFT configurations, the bowling--pin (BP) $5\alpha$ cluster model fitted to NLEFT, the $^{16}\mathrm{O}+\alpha$ cluster configuration, and the spherical 3pf Woods--Saxon distribution.

For all centralities, the 3pf Woods--Saxon case yields the largest
$\sqrt{\langle \mathcal{P}^{2}\rangle}$.
Introducing $\alpha$ clustering systematically suppresses the
fluctuation of the net spin.
This suppression is strongest for the BP and NLEFT configurations,
which both realize a compact $5\alpha$ geometry with pronounced local
spin cancellation.
The $^{16}\mathrm{O}+\alpha$ configuration shows an intermediate behavior:
its $\sqrt{\langle \mathcal{P}^{2}\rangle}$ curve lies below the
3pf Woods--Saxon reference, but above the BP and NLEFT results, indicating
that the detached $\alpha$ cluster at larger radius generates weaker
short-range spin anti-correlations than the BP core with $5\alpha$ clusters.
These trends become more transparent after removing the trivial
finite-size effect.

\begin{figure}[t]
\begin{center}
\includegraphics[width=0.95\linewidth]{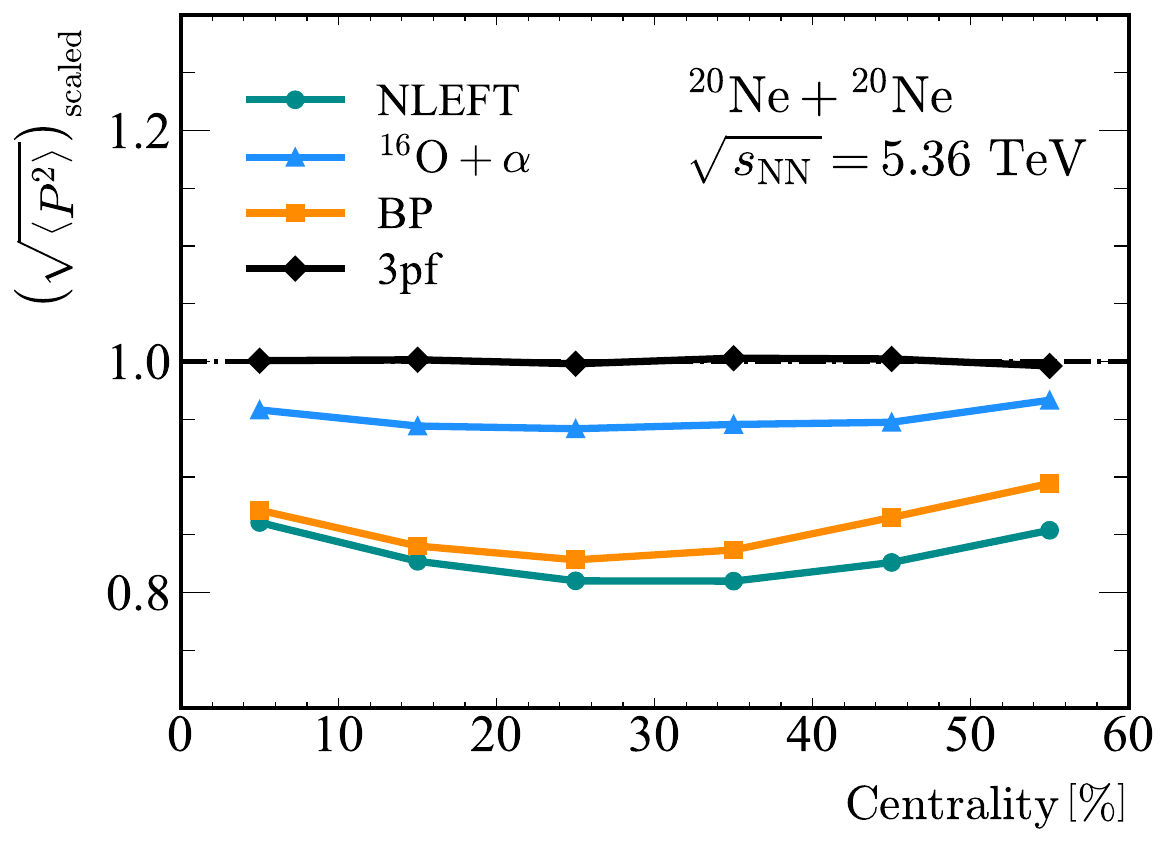}
\end{center}
\caption{(Color online) Centrality dependence of the scaled standard deviation of 
the polarization parameter $\left(\sqrt{\langle \mathcal{P}^{2}\rangle}\right)_{\mathrm{scaled}}$ 
in $^{20}\mathrm{Ne}+{}^{20}\mathrm{Ne}$ collisions at $\sqrt{s_{\mathrm{NN}}}=5.36~\mathrm{TeV}$. 
The teal circles, blue triangles, orange squares, and black diamonds correspond to the
NLEFT, $^{16}\mathrm{O}+\alpha$, BP, and spherical 3pf Woods--Saxon configurations, 
respectively. The horizontal dashed line indicates the reference value of unity.}
\label{fig:std_p2_scaled_Neon}
\end{figure}
In Fig.~\ref{fig:std_p2_scaled_Neon} we show the scaled standard
deviation $\left(\sqrt{\langle \mathcal{P}^{2}\rangle}\right)_{\text{scaled}}$.
The 3pf Woods--Saxon curve remains close to unity.
In contrast, all clusterized configurations yield
$\left(\sqrt{\langle \mathcal{P}^{2}\rangle}\right)_{\text{scaled}}<1$ over the
whole centrality range, indicating additional spin anti-correlations
beyond the theory-defined, $J=0$-conditioned null baseline.

The suppression is most pronounced at intermediate centralities where
$N_{\text{part}}$ is large enough to sample several clusters but not so
large that the entire nucleus is fully overlapped.
We observe a characteristic non-monotonic behavior:
$\left(\sqrt{\langle \mathcal{P}^{2}\rangle}\right)_{\text{scaled}}$ first
decreases from central to mid-central collisions, reaches a minimum,
and then increases again towards peripheral events.
Physically, this pattern reflects the interplay between the number of
independent $\alpha$ clusters participating in the collision and the
efficiency of local spin cancellation within each cluster.
The NLEFT and BP curves nearly coincide, indicating that the fitted
BP geometry captures most of the spin-structure information contained
in the NLEFT configurations for $^{20}\mathrm{Ne}$.
By contrast, the $^{16}\mathrm{O}+\alpha$ configuration remains closer
to the 3pf Woods--Saxon reference, which suggests that measurements of
$\left(\sqrt{\langle \mathcal{P}^{2}\rangle}\right)_{\text{scaled}}$ in
$^{20}\mathrm{Ne}+{}^{20}\mathrm{Ne}$ collisions can discriminate
between a genuine $5\alpha$ bowling--pin structure and a more loosely
bound $^{16}\mathrm{O}+\alpha$ configuration ~\cite{Li:2025NeNe,Yamaguchi:2023,Bijker:2021Ne20,Adachi:2021fivealpha,Kawabata2024FiveAlphaCondensate20Ne}.

\subsection{\texorpdfstring{$^{16}\mathrm{O}+{}^{16}\mathrm{O}$}{O+O} collisions}

\begin{figure}[t]
\begin{center}
\includegraphics[width=0.95\linewidth]{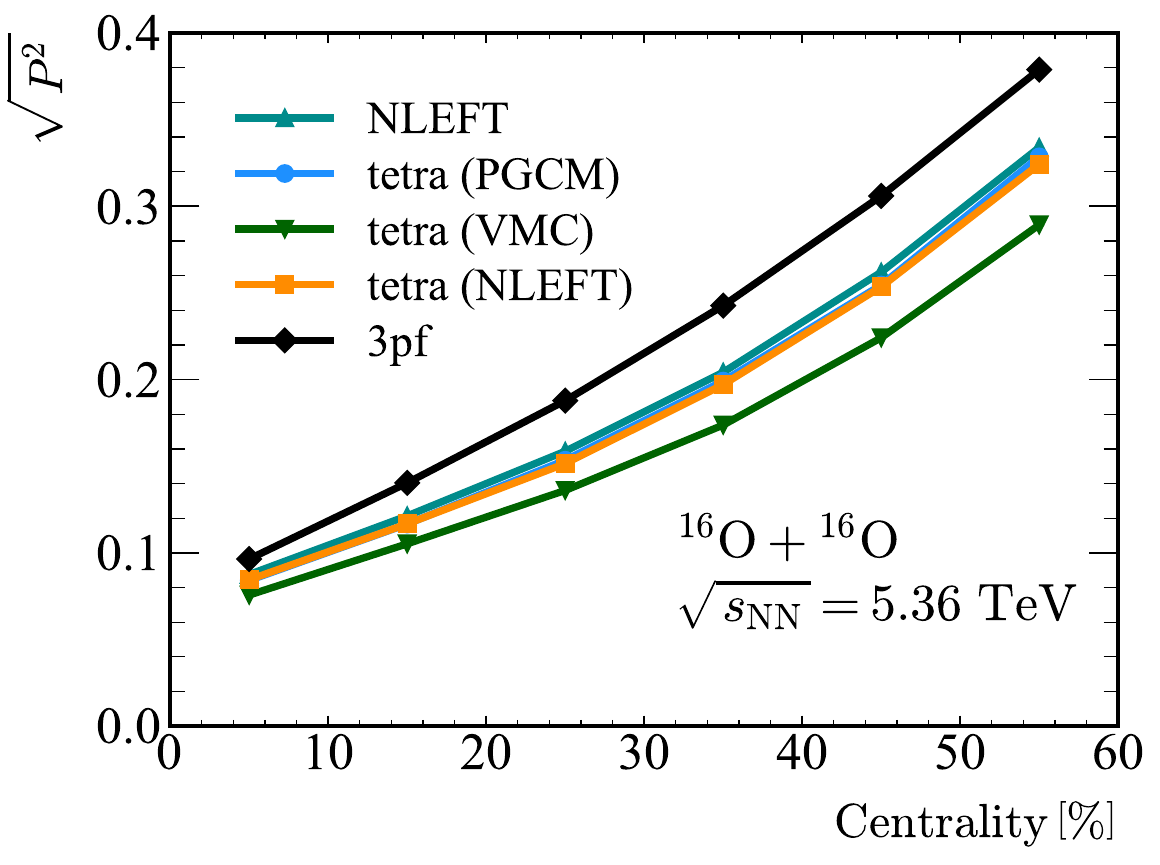}
\end{center}
\caption{(Color online) Standard deviation of the polarization parameter 
$\sqrt{\langle \mathcal{P}^{2}\rangle}$ as a function of centrality in 
$^{16}\mathrm{O}+{}^{16}\mathrm{O}$ collisions at $\sqrt{s_{\mathrm{NN}}}=5.36~\mathrm{TeV}$.
Teal triangles show the \textit{ab initio} NLEFT result. ``tetra (PGCM)'', ``tetra (VMC)'', 
and ``tetra (NLEFT)'' (blue circles, green downward triangles, and orange squares) 
are tetrahedral $4\alpha$ configurations fitted to PGCM, VMC, and NLEFT 
\textit{ab initio} data, respectively, while ``3pf'' (black diamonds) denotes a spherical 3pf 
Woods--Saxon distribution.}
\label{fig:std_p2_Oxygen}
\end{figure}

We now turn to $^{16}\mathrm{O}+{}^{16}\mathrm{O}$ collisions.
Fig.~\ref{fig:std_p2_Oxygen} shows $\sqrt{\langle \mathcal{P}^{2}\rangle}$ for
the NLEFT configurations, for tetrahedral $4\alpha$ cluster models
fitted to NLEFT, PGCM, and VMC densities, and for the spherical
3pf Woods--Saxon reference.

As in $^{20}\mathrm{Ne}+{}^{20}\mathrm{Ne}$ collisions, the 3pf Woods--Saxon distribution yields the largest
spin fluctuations, while the presence of a tetrahedral $4\alpha$
geometry leads to a visible suppression of $\sqrt{\langle \mathcal{P}^{2}\rangle}$. The NLEFT results are close to the tetrahedral configurations fitted to NLEFT and PGCM densities.
The VMC-based tetrahedron is slightly different, reflecting its more compact $\alpha$ clusters (smaller $r_L$) and modified intercluster distance $\ell_c$.
This shows that $\sqrt{\langle \mathcal{P}^{2}\rangle}$ is sensitive not only
to the existence of clustering, but also to the detailed geometry of
the clusters encoded in the ratio $r_L/\ell_c$~\cite{Shafi:2025feq,Wang:2024compact}.

\begin{figure}[t]
\begin{center}
\includegraphics[width=0.95\linewidth]{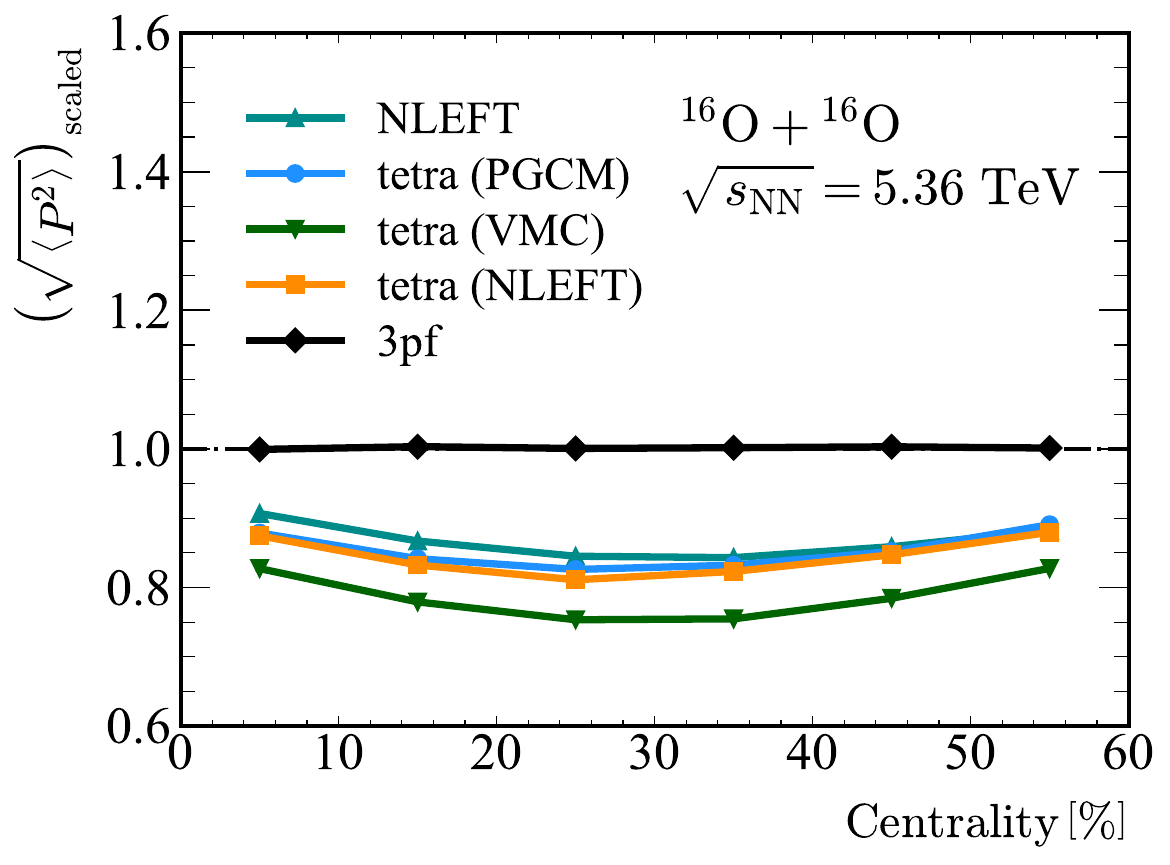}
\end{center}
\caption{(Color online) Centrality dependence of the scaled standard deviation of 
the polarization parameter $\left(\sqrt{\langle \mathcal{P}^{2}\rangle}\right)_{\text{scaled}}$ 
in $^{16}\mathrm{O}+{}^{16}\mathrm{O}$ collisions at $\sqrt{s_{\mathrm{NN}}}=5.36~\mathrm{TeV}$. 
Teal triangles show the \textit{ab initio} NLEFT result. The curves labeled ``tetra (PGCM)'', 
``tetra (VMC)'', and ``tetra (NLEFT)'' (blue circles, green downward triangles, and 
orange squares) correspond to tetrahedral $4\alpha$ configurations fitted to 
PGCM, VMC, and NLEFT \textit{ab initio} data, respectively, while ``3pf'' (black diamonds) 
denotes a spherical 3pf Woods--Saxon distribution. The horizontal dashed line marks the reference 
value of unity.}
\label{fig:std_p2_scaled_Oxygen}
\end{figure}
Fig.~\ref{fig:std_p2_scaled_Oxygen} presents the scaled quantity,
$\left(\sqrt{\langle \mathcal{P}^{2}\rangle}\right)_{\text{scaled}}$, in $^{16}\mathrm{O}+{}^{16}\mathrm{O}$ collisions at $\sqrt{s_{\mathrm{NN}}}=5.36~\mathrm{TeV}$.
Again, the 3pf Woods--Saxon curve stays close to unity, whereas all clusterized configurations give values below unity, with a centrality
dependence similar to that observed in $^{20}\mathrm{Ne}+{}^{20}\mathrm{Ne}$.
The suppression is strongest at mid-central collisions, where the
spectator region is still sizable but a large fraction of the
tetrahedral structure participates in the interaction.
Moreover, the three tetrahedral fits exhibit a clear ordering:
the configuration with the most compact clusters (smallest $r_L$ for
a given $\ell_c$) shows the largest deviation from the 3pf Woods--Saxon
reference.
This shows that
$\left(\sqrt{\langle \mathcal{P}^{2}\rangle}\right)_{\text{scaled}}$ is a
useful diagnostic of the underlying cluster spin structure, sensitive
to subtle differences between modern \textit{ab initio} descriptions of
$^{16}\mathrm{O}$~\cite{Zhang:2024vkh}.

\subsection{System-size ratios and experimental prospects}

\begin{figure}[t]
\begin{center}
\includegraphics[width=0.95\linewidth]{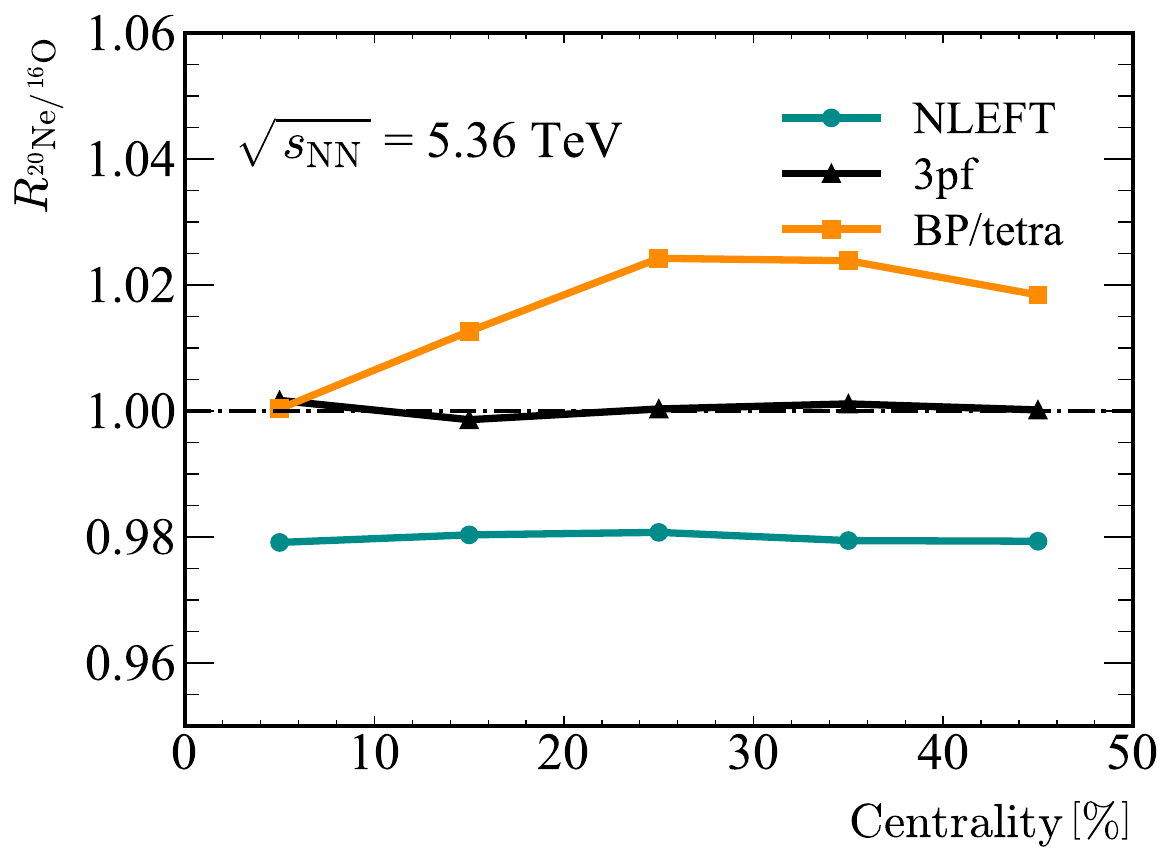}
\end{center}
\caption{(Color online) Centrality dependence of the ratio 
$R_{^{20}\mathrm{Ne}/^{16}\mathrm{O}}$ of the scaled standard deviation of the 
polarization parameter $\left(\sqrt{\langle \mathcal{P}^{2}\rangle}\right)_{\text{scaled}}$ 
between $^{20}\mathrm{Ne}+{}^{20}\mathrm{Ne}$ and $^{16}\mathrm{O}+{}^{16}\mathrm{O}$ 
collisions at $\sqrt{s_{\mathrm{NN}}}=5.36~\mathrm{TeV}$. Teal circles show the 
ratio obtained from \textit{ab initio} NLEFT initial states, black triangles the ratio for 
spherical 3pf Woods--Saxon nuclei, and orange squares the ratio for BP-shaped 
$^{20}\mathrm{Ne}$ and tetrahedral $^{16}\mathrm{O}$ configurations fitted to 
NLEFT data.}
\label{fig:Neon_Oxygen_ratio}
\end{figure}

While the scaled fluctuation already removes most of the trivial
finite-size dependence at the model-comparison level, an experimental analysis based on
final-state spin-correlation observables in a single system may still suffer from residual systematic uncertainties and
dynamical attenuation of the spin signal during the hydrodynamic evolution.
To further suppress such effects, we consider the system-size ratio 
$R_{^{20}\mathrm{Ne}/^{16}\mathrm{O}}$, defined in Eq.~\eqref{eq:R_ratio_def}, 
as presented in Fig.~\ref{fig:Neon_Oxygen_ratio}.

For spherical 3pf Woods--Saxon nuclei, the ratio stays very close to
unity over the entire centrality range.
Deviations of $R_{^{20}\mathrm{Ne}/^{16}\mathrm{O}}$ from unity in the
clustered cases therefore reflect differences in the spin structure of
the two nuclei.

The \textit{ab initio} NLEFT configurations yield a ratio that is
slightly below one, deviating by a few percent from the 3pf Woods--Saxon
reference.
In contrast, when both nuclei are modeled by simple cluster
configurations fitted to NLEFT densities (BP-shaped $^{20}\mathrm{Ne}$
and tetrahedral $^{16}\mathrm{O}$), the ratio lies slightly above unity
and exhibits a weak increase with centrality, approaching values of the
order of $1.02$ in the most peripheral bins.
The opposite sign of the deviation in these two NLEFT-based scenarios
indicates that the minimal $\alpha$-cluster models do not capture all
the spin-structure information present in the full \textit{ab initio}
wave functions.
In other words, the \textit{ab initio} NLEFT configurations encode additional, more intricate spin--spin correlations beyond those associated with idealized $\alpha$ clusters, possibly including correlations between different clusters that are not captured in the minimal cluster parametrizations.

From an experimental perspective, the pattern in
Fig.~\ref{fig:Neon_Oxygen_ratio} is particularly attractive.
The 3pf Woods--Saxon ratio provides a useful reference close to unity,
while realistic clustered nuclei produce deviations of order
$2\%$ that are roughly constant or slowly varying with
centrality.
Such percent-level differences are similar in magnitude to the
initial-spin signals estimated for heavy systems in
Ref.~\cite{Giacalone:2025}, and may be within reach of high-statistics
$^{16}\mathrm{O}+{}^{16}\mathrm{O}$ and $^{20}\mathrm{Ne}+{}^{20}\mathrm{Ne}$
runs at the LHC.
Under the working assumption that residual system-size and
hydrodynamic effects are largely common-mode between the two collision systems,
this ratio is expected to mitigate a significant part of those contributions,
thereby enhancing the sensitivity to differences in their intrinsic spin structures.
A combined measurement of final-state spin-correlation observables in
$^{20}\mathrm{Ne}+{}^{20}\mathrm{Ne}$ and $^{16}\mathrm{O}+{}^{16}\mathrm{O}$
collisions, together with comparison to the corresponding scaled quantities and their ratio, would thus provide a new, spin-based
handle on clustering in light nuclei that is complementary to
traditional low-energy observables.

\section{Summary and outlook}
\label{sec:summary}

In this work we have explored, for the first time, how the spin structure
of light $\alpha$-clustered nuclei affects initial spin fluctuations in
ultra-relativistic nucleus--nucleus collisions within the theoretical
framework that relates initial spin fluctuations to final-state
$\Lambda$-pair spin correlations~\cite{Giacalone:2025}.
In particular, we focused on $^{16}\mathrm{O}$ and $^{20}\mathrm{Ne}$,
where the ground states are dominated by $2p$--$2n$ $\alpha$ clusters in
which spin and isospin are locally cancelled, leading to pronounced
short-range spin--spin anti-correlations.
Our aim was to quantify how such microscopic correlations manifest
themselves in the event-by-event fluctuations of the net spin
polarization of the fireball.

As nuclear-structure input we employed \textit{ab initio} configurations
from Nuclear Lattice Effective Field Theory (NLEFT), which provide for
each event the full set of nucleon coordinates and spin quantum numbers,
together with several phenomenological $\alpha$-cluster geometries and
spherical 3pf Woods--Saxon distributions without
intrinsic spin--spin correlations.
For each of these models we generated Monte-Carlo samples of
$^{16}\mathrm{O}+{}^{16}\mathrm{O}$ and $^{20}\mathrm{Ne}+{}^{20}\mathrm{Ne}$
collisions at $\sqrt{s_{\mathrm{NN}}}=5.36$~TeV using the \trento{}
framework, and evaluated both the standard deviation
$\sqrt{\langle \mathcal{P}^2\rangle}$ of the initial polarization and its scaled
version defined using the theory-defined, $J=0$-conditioned null baseline.
We find that 3pf Woods--Saxon nuclei exhibit the largest
$\sqrt{\langle \mathcal{P}^2\rangle}$, whereas clusterized and NLEFT
configurations show a significant suppression due to local spin--spin
anti-correlations inside the $\alpha$ clusters.
After scaling, all clustered configurations yield
$\left(\sqrt{\langle \mathcal{P}^2\rangle}\right)_{\text{scaled}}<1$ with a
characteristic non-monotonic centrality dependence.
The NLEFT results are closely reproduced by a bowling--pin $5\alpha$
geometry for $^{20}\mathrm{Ne}$ and by a tetrahedral $4\alpha$
configuration for $^{16}\mathrm{O}$ fitted to NLEFT densities, indicating
that these simple models capture the main features of the underlying
cluster spin structure.

Beyond this overall suppression, the scaled fluctuations are sensitive
to the detailed cluster geometry.
In $^{16}\mathrm{O}+{}^{16}\mathrm{O}$ collisions, different tetrahedral
configurations fitted to NLEFT, PGCM, and VMC densities exhibit a clear
ordering that reflects the compactness of the clusters, encoded in the
ratio $r_L/\ell_c$.
In $^{20}\mathrm{Ne}+{}^{20}\mathrm{Ne}$ collisions,
$\left(\sqrt{\langle \mathcal{P}^2\rangle}\right)_{\text{scaled}}$ distinguishes
between a compact bowling--pin $5\alpha$ structure and a more
loosely bound $^{16}\mathrm{O}+\alpha$ configuration.
We further proposed the system-size ratio of scaled fluctuations,
$R_{^{20}\mathrm{Ne}/^{16}\mathrm{O}}$, as a more robust discriminator.
For 3pf Woods--Saxon nuclei this ratio stays very close to unity,
whereas clustered nuclei produce percent-level deviations:
for bowling--pin $^{20}\mathrm{Ne}$ and tetrahedral $^{16}\mathrm{O}$
fitted to NLEFT densities the ratio increases with centrality and
reaches values of about $1.02$ in more peripheral bins, while
for full NLEFT configurations it remains slightly below one, around
$0.98$.
The opposite sign of the deviation suggests that NLEFT contains more
intricate spin--spin correlations than those encoded in minimal
$\alpha$-cluster models, underscoring the potential of
$R_{^{20}\mathrm{Ne}/^{16}\mathrm{O}}$ as a sensitive probe of nuclear
spin structure.

In summary, percent-level deviations of $R_{^{20}\mathrm{Ne}/^{16}\mathrm{O}}$ from unity would provide evidence for nontrivial spin structures associated with $\alpha$ clustering in light nuclei.

Several extensions of the present study are worth pursuing.
First, the initial spin density $\mathcal{S}(\mathbf{x})$ constructed
here should be evolved within full relativistic spin hydrodynamics~\cite{Huang:2025IntroSpinHydro,Drogosz:2024HybridSpinHydro} in
order to quantify the dynamical attenuation of spin fluctuations from
the initial state to the final-state $\Lambda$ polarization.
More realistic initial spin models, including spin transport during the
early nucleon--nucleon scatterings and a construction of initial spin
densities that is more closely tied to quantum-mechanical principles~\cite{Ke:2025tyv},
would further improve the predictive power.
Second, it will be interesting to extend the present analysis to other
nuclear systems, such as isotopic pairs like $^{96}\mathrm{Ru}$ and
$^{96}\mathrm{Zr}$~\cite{Zhang:2021kxj,Xi2025MomentumCorrelationRuZr}, which may help disentangle proton--proton,
neutron--neutron, and proton--neutron spin correlations.
Third, since each $\alpha$ cluster is an approximate spin--isospin
singlet, the same local cancellation mechanism applies equally to
isospin. Cluster-induced initial-state isospin correlations may
partially survive the fireball evolution and leave imprints on
light-nucleus yields and yield ratios~\cite{Bleicher2024NucleosynthesisLittleBang,Feng2026LightNucleiCorrelations},
particularly in light systems where the shorter fireball lifetime
favors the survival of such correlations.
Finally, additional ab initio inputs for light and medium-mass nuclei
will allow one to better constrain first-principles descriptions of
nuclear spin structure; optimized generator-coordinate basis construction for microscopic cluster models may also provide useful complementary inputs in future applications~\cite{Liu2025OptimizingBasisGCM}. The percent-level differences
predicted here for $R_{20{\rm Ne}/16{\rm O}}$ may be within reach of future
high-luminosity $^{16}$O and $^{20}$Ne runs at the LHC. More broadly, our results motivate a spin-based comparison program in light-ion collisions: measure final-state spin-correlation observables in multiple systems, and confront them with theory-guided normalized quantities that enhance sensitivity to intrinsic nuclear spin structure.

\begin{acknowledgments}
This paper is partly dedicated to commemorating Prof. Xu Cai (IOPP, CCNU), who made outstanding contributions to the research on relativistic heavy ion collisions.
We are grateful to Ulf-G. Mei{\ss}ner, Dean Lee, and Shihang Shen for providing the \textit{ab initio} nuclear configurations from Nuclear Lattice Effective Field Theory (NLEFT) calculations. 
We especially thank Shihang Shen for preparing the specific datasets for $^{16}\mathrm{O}$ and $^{20}\mathrm{Ne}$ with spin information and for valuable discussions on their usage. 
We thank Giuliano Giacalone for inspiring this study and for enlightening discussions regarding the theoretical framework.
We also appreciate helpful comments from Weiyao Ke and Dean Lee.
This work was supported by the National Natural Science Foundation of China (NSFC) under Grants No.~12305138 and No.~12535010, the Guangdong Major Project of Basic and Applied Basic Research under Grant No.~2020B0301030008. 
X.F. acknowledges support from the China Scholarship Council (CSC).
\end{acknowledgments}

\bibliographystyle{apsrev4-1}

\bibliography{inispin}

\appendix

\section{Spin-hydrodynamic conventions and connection to the initial state}
\label{app:spin_hydro}

This appendix summarizes the spin-hydrodynamic conventions and working
assumptions used in Sec.~\ref{subsec:spin_model}. Its purpose is to
clarify how the spin tensor $S^{\lambda,\mu\nu}$, the spin-density
four-vector $S^\mu$, and the initial scalar density
$\mathcal{S}({\bf x})$ are connected in the present work.
It is not intended as a general review of relativistic spin
hydrodynamics.

The equations of motion are given by the conservation of
energy--momentum and total angular momentum~\cite{Florkowski:2018RelHydroSpin,Hongo:2021SpinHydroTorsion},
\begin{equation}
  \partial_\mu T^{\mu\nu} = 0,
  \qquad
  \partial_\lambda J^{\lambda,\mu\nu} = 0,
  \label{eq:hydro_eom_app}
\end{equation}
where $T^{\mu\nu}$ is the energy--momentum tensor and
$J^{\lambda,\mu\nu}$ the total angular-momentum tensor.
The latter can be decomposed into orbital and spin contributions,
\begin{equation}
  J^{\lambda,\mu\nu}
  = L^{\lambda,\mu\nu} + S^{\lambda,\mu\nu},
  \qquad
  L^{\lambda,\mu\nu}
  = x^\mu T^{\lambda\nu} - x^\nu T^{\lambda\mu}.
\end{equation}
Inserting this decomposition into Eq.~\eqref{eq:hydro_eom_app} and using
$\partial_\mu T^{\mu\nu}=0$ one obtains
\begin{equation}
  \partial_\lambda S^{\lambda,\mu\nu}
  = T^{\nu\mu} - T^{\mu\nu} ,
  \label{eq:spin_noncons_app}
\end{equation}
which shows that the spin tensor $S^{\lambda,\mu\nu}$ is exactly
conserved only if the energy--momentum tensor is symmetric.
Different pseudo-gauge choices correspond to different decompositions
of $T^{\mu\nu}$ and $S^{\lambda,\mu\nu}$, but leave the total charges
associated with $J^{\lambda,\mu\nu}$ invariant~\cite{Weickgenannt:2022PseudoGauge,Hongo:2021SpinHydroTorsion}.
Accordingly, Eq.~\eqref{eq:hydro_EOM_main} in the main text should be
understood as the working form adopted here, corresponding to
pseudo-gauges in which the relevant $T^{\mu\nu}$ is symmetric.

It is convenient to introduce the spin-density tensor
\begin{equation}
  S^{\mu\nu} \equiv u_\lambda S^{\lambda,\mu\nu},
\end{equation}
where $u^\mu$ is the fluid four-velocity.
From $S^{\mu\nu}$ one defines the spin-density four-vector
\begin{equation}
  S^\mu \equiv
  \frac{1}{2}\,\epsilon^{\mu\nu\alpha\beta}\,
  S_{\nu\alpha}\,u_\beta ,
  \qquad
  S^\mu u_\mu = 0,
  \label{eq:Smu_def_app}
\end{equation}
which contains the three independent spatial components in the local
rest frame.

In the initial state of a high-energy nuclear collision we assume boost
invariance around midrapidity and vanishing transverse flow, so that
$u^\mu=(1,0,0,0)$ at $\tau\to 0^+$.
Following Ref.~\cite{Giacalone:2025}, only the spatial components of
$S^\mu$ are taken to be nonzero initially and are parameterized as
$S^i = n^i S$, with a unit vector $n^i$ defining the random orientation
of the event-wise polarization and a scalar amplitude $S$, see
Eq.~\eqref{eq:S_vec}.
The corresponding initial spin density per unit rapidity at
midrapidity, Eq.~\eqref{eq:S_rapidity}, is implemented in our model
through the smeared participant sum $\mathcal{S}({\bf x})$ in
Eq.~\eqref{eq:S_field}.
In this sense, $\mathcal{S}({\bf x})$ provides the operational mapping
between the initial-state participant spin distribution and the
spin-density field used as input to spin-hydrodynamic evolution.

The role of this appendix is thus only to make explicit the conventions
underlying the use of Eqs.~\eqref{eq:S_vec}--\eqref{eq:S_field} and
Eq.~\eqref{eq:hydro_EOM_main} in the main text.
For broader discussions of relativistic spin hydrodynamics, see
Refs.~\cite{Florkowski:2018RelHydroSpin,Florkowski:2019RelSpinHydro,Speranza:2021SpinHydroReview,Giacalone:2025}.

\section{Polarization variance and scaled spin fluctuations}
\label{app:P2_spin}

In this appendix we collect the relations between the event-by-event
polarization variance, spin--spin correlations in the colliding nuclei,
and the baseline and scaled quantities used in the main text.

The event-wise polarization is defined in Eq.~\eqref{eq:P_def} as
\begin{equation}
  \mathcal{P}
  = \frac{1}{\Npart}
    \left(
      \sum_{i=1}^{\NA} s_{i,A}
      +
      \sum_{j=1}^{\NB} s_{j,B}
    \right),
  \qquad
  \Npart = \NA + \NB ,
\end{equation}
where $s_{i,A}=\pm1$ and $s_{j,B}=\pm1$ denote spin projections of
participants from nuclei $A$ and $B$, respectively.
Here $\NA$ and $\NB$ denote the numbers of participants from nuclei $A$
and $B$ in a given event, with $\Npart=\NA+\NB$; these quantities
fluctuate event by event within a centrality class.
For unpolarized colliding ions the one-body spin density is symmetric,
so that $\langle s_{i,A}\rangle=\langle s_{j,B}\rangle=0$ and
$\langle\mathcal{P}\rangle=0$.

The variance of $\mathcal{P}$ in a given centrality class is
\begin{equation}
\langle\mathcal{P}^2\rangle
\equiv
\Bigg\langle
\frac{1}{\Npart^{2}}
\left(
  \sum_{i=1}^{\NA} s_{i,A}
  +
  \sum_{j=1}^{\NB} s_{j,B}
\right)^2
\Bigg\rangle .
\end{equation}

Because spin degrees of freedom in the two nuclei are
independent, the mixed terms factorize and vanish,
\begin{equation}
  \langle s_{i,A} s_{j,B} \rangle
  = \langle s_{i,A} \rangle
    \langle s_{j,B} \rangle
  = 0 ,
\end{equation}
and one obtains
\begin{equation}
\langle\mathcal{P}^2\rangle
=
\Bigg\langle
\frac{1}{\Npart^{2}}
\left[
  \sum_{i,k=1}^{\NA} s_{i,A}s_{k,A}
  +
  \sum_{j,l=1}^{\NB} s_{j,B}s_{l,B}
\right]
\Bigg\rangle .
\label{eq:P2_decomp_app}
\end{equation}

In the following, angle brackets acting on spin products such as
$\langle s_{i,A}s_{k,A}\rangle$ denote the expectation value over spin degrees of freedom
for a given participant configuration, while the outer $\langle\cdots\rangle$
denotes the average over events in the centrality class.

Consider, for definiteness, the contribution from nucleus $A$.
It contains $\NA$ diagonal terms with $i=k$ and
$\NA(\NA-1)$ off-diagonal terms with $i\neq k$,
\begin{equation}
\begin{split}
  \sum_{i,k=1}^{\NA}
    \langle s_{i,A}s_{k,A} \rangle
  &=
  \NA
  \sum_{s=\pm1}
    \rho_A^{(1)}(s)\,s^2 \\
  &\quad +
  \NA(\NA-1)
  \sum_{s_1,s_2=\pm1}
    \rho_A^{(2)}(s_1,s_2)\,s_1 s_2 ,
  \raisetag{52pt}
\end{split}
  \label{eq:AA_corr_app}
\end{equation}

where $\rho_A^{(1)}(s)$ is the one-body spin density of participants
from nucleus $A$, and $\rho_A^{(2)}(s_1,s_2)$ is the
two-body spin density for nucleus $A$.
An analogous expression holds for nucleus $B$.
Inserting Eq.~\eqref{eq:AA_corr_app} (and its $B$ counterpart) into
Eq.~\eqref{eq:P2_decomp_app} yields
\begin{equation}
\begin{split}
\langle\mathcal{P}^2\rangle
=
\Bigg\langle
\frac{1}{\Npart^{2}}
\bigg[
  &\NA \sum_{s} \rho_A^{(1)}(s)\,s^2 \\
  &+ \NA(\NA-1) \sum_{s_1,s_2} \rho_A^{(2)}(s_1,s_2)\,s_1 s_2 \\
  &+ \NB \sum_{s} \rho_B^{(1)}(s)\,s^2 \\
  &+ \NB(\NB-1) \sum_{s_1,s_2} \rho_B^{(2)}(s_1,s_2)\,s_1 s_2
\bigg]
\Bigg\rangle .
\end{split}
\label{eq:P2_full_app}
\end{equation}

For collisions of identical nuclei one has
$\rho_A^{(n)}=\rho_B^{(n)}$ and Eq.~\eqref{eq:P2_full_app} reduces, up
to an overall normalization factor, to the schematic relation
\eqref{eq:P2_rho_main} given in the main text.
The dependence on the one-body densities $\rho^{(1)}$ reflects the
trivial variance of independent spins, while the terms containing
$\rho^{(2)}$ encode genuine two-body spin--spin correlations, such as
those induced by $\alpha$ clustering.

To isolate nontrivial spin structures, we introduce a null baseline that enforces the global $J=0$ constraint and otherwise assumes no additional intrinsic spin correlations.
Consider a single nucleus with mass number $A_{\text{mass}}$ and spin variables $s_i=\pm1$.
In this simplified spin representation, the $J=0$ condition is implemented as
\begin{equation}
  \sum_{i=1}^{A_{\text{mass}}} s_i = 0 .
\end{equation}
Squaring this relation and taking an expectation value gives
\begin{equation}
\begin{aligned}
  0
  &= \left\langle
      \left(\sum_{i=1}^{A_{\text{mass}}} s_i\right)^2
    \right\rangle \\
  &= A_{\text{mass}}\langle s_i^2\rangle
    + A_{\text{mass}}(A_{\text{mass}}-1)
      \langle s_i s_j \rangle_{\text{base}} ,
\end{aligned}
\end{equation}
where $\langle s_i^2\rangle=1$ and, within the null-baseline ensemble, permutation symmetry implies that all distinct pairs
$(i\neq j)$ have the same correlation
$\langle s_i s_j \rangle_{\text{base}}$.
Solving for this pair correlation one finds
\begin{equation}
  c=\langle s_i s_j \rangle_{\text{base}}
  = -\frac{1}{A_{\text{mass}}-1} ,
  \label{eq:base_pair_app}
\end{equation}
which is the ``trivial'' negative background correlation quoted in the
main text.

Now consider a collision between nuclei $A$ and $B$ and restrict
attention to participants.
Assuming that the pair correlation in the null baseline between any two nucleons
from the same nucleus is given by the value in
Eq.~\eqref{eq:base_pair_app}, independent of whether the nucleons participate in the reaction, the
baseline variance of the polarization follows from
Eq.~\eqref{eq:P2_decomp_app}.
For each nucleus the diagonal and off-diagonal contributions read
\begin{equation}
  \sum_{i,k=1}^{\NA}
    \langle s_{i,A}s_{k,A} \rangle_{\text{base}}
  =
  \NA
  + \NA(\NA-1)c ,
\end{equation}
the resulting baseline variance is
\begin{equation}
\begin{aligned}
  \langle\mathcal{P}^2\rangle_{\text{base}}
  &=
  \Bigg\langle
  \frac{1}{\Npart^{2}}
  \Big[
    \NA + \NB \\
  &\quad\; + c\,\NA(\NA-1)
           + c\,\NB(\NB-1)
  \Big]
  \Bigg\rangle \\
  &=
\Bigg\langle
\frac{1}{\Npart^{2}}
\Big[
  \Npart
  + c \sum_{X=A,B} N_{\text{part}}^{X}\big(N_{\text{part}}^{X}-1\big)
\Big]
\Bigg\rangle ,
\end{aligned}
  \label{eq:P2_base_app}
\end{equation}

which is equivalent to Eq.~\eqref{eq:P2_base_main} used in
Sec.~\ref{subsec:observables}.
By construction, this quantity is evaluated in the null baseline conditioned on the global $J=0$ constraint,
and therefore contains no information on additional intrinsic spin correlations.

Since $\langle\mathcal{P}\rangle=0$, the standard deviation of the
polarization is simply
\begin{equation}
  \text{std}(\mathcal{P})
  = \sqrt{\langle\mathcal{P}^2\rangle} .
\end{equation}
Accordingly, the scaled standard deviation plotted in
Sec.~\ref{sec:results} is defined as
\begin{equation}
  \bigl(\sqrt{\langle\mathcal{P}^2\rangle}\bigr)_{\text{scaled}}
  =
  \frac{\sqrt{\langle\mathcal{P}^2\rangle}}
       {\sqrt{\langle\mathcal{P}^2\rangle_{\text{base}}}} ,
  \label{eq:P2_scaled_def_app}
\end{equation}
which equals unity for the null baseline conditioned on $J=0$ and deviates
from unity only in the presence of additional intrinsic spin correlations beyond this baseline.

\section{Processing of NLEFT configurations}
\label{app:NLEFT}
This appendix summarizes how the \textit{ab initio} NLEFT pinhole
configurations are processed and used as nuclear-structure input for the
model simulations of Sec.~\ref{sec:results}.

\paragraph{(1) NLEFT configurations.}
For each nucleus considered in this work, NLEFT provides an ensemble of
ground-state samples generated on a periodic spatial lattice with chiral
effective interactions.
Each configuration specifies the occupied lattice sites of all nucleons,
together with their spin and isospin quantum numbers from the pinhole
algorithm, so that many-body correlations in both coordinate and
spin--isospin space are preserved.
In practice, the stored coordinates are integer lattice indices
$(n_x,n_y,n_z)$ with $n_\alpha\in\{0,\dots,L-1\}$, and the spin quantum
numbers are given as an $s_z$-type projection along a fixed quantization
axis in the configuration.

\paragraph{(2) Sampling of nucleon positions.}
To obtain continuous nucleon coordinates from the lattice sites, we
sample independently a uniform sub-cell displacement
$\delta n_\alpha\in[-1/2,\,1/2]$ for each nucleon and each Cartesian
direction $\alpha=x,y,z$, and set $n_\alpha' = n_\alpha+\delta n_\alpha$.
If $n_\alpha'$ falls outside the interval $[0,L)$, it is wrapped back
into the box according to periodic boundary conditions.
We then shift each configuration to its center-of-mass frame using the
minimum-image convention and convert to physical units using the lattice
spacing $a$.
The resulting nucleon positions ${\bf x}_i$, together with the spin and
isospin labels, are passed to \trento{}, where the entire configuration is
randomly rotated event by event to sample isotropic orientations of the
intrinsic nuclear shape relative to the beam axis.

\paragraph{(3) Spin labels and quantization axis.}
The NLEFT spin quantum numbers are stored as projections along a fixed
quantization axis (an $s_z$-type projection) in each configuration.
Since \trento{} randomly rotates the full configuration event by event,
this quantization direction is effectively sampled isotropically in the
laboratory frame.
We therefore map the stored spin eigenvalues onto two-valued labels
$s_i=\pm1$ and use them as the spin inputs entering scalar observables
such as $s_i s_j$ and $\mathcal{P}$, thereby preserving the local
spin--spin correlations encoded in the NLEFT ensemble.
The information required to re-project spins onto an independently chosen
axis is not available from the stored configurations; any residual
dependence on the original quantization convention, if present, should be
regarded as a potential systematic uncertainty that we do not quantify
in this work.

The two-nucleon spin-correlation function shown in Fig.~\ref{fig:sisj}
is obtained by averaging the product $s_i s_j$ over all pairs of nucleons
in the NLEFT ensemble and binning the results as a function of their
relative distance $\delta r = |{\bf x}_i-{\bf x}_j|$.

\end{document}